\documentclass[twocolumn]{aastex631}

\usepackage{graphicx}
\usepackage{amsmath}
\usepackage{pifont}
\usepackage[caption=false]{subfig}

\newcommand{\nhi}{$N_{\rm HI}$}
\newcommand{\nhii}{$N_{\rm HII}$}

\newcommand{\cmark}{\ding{51}}
\newcommand{\xmark}{\ding{55}}

\begin{document}

\title{The Metallicity Mapping of the Ionized Diffuse Gas at the Milky Way Disk-halo Interface}

%\correspondingauthor{August Muench}
%\email{greg.schwarz@aas.org, gus.muench@aas.org}

\author[0000-0002-9040-672X]{Bo-Eun Choi}
\affiliation{Department of Astronomy, University of Washington, 
Seattle, WA 98195, USA}

\author[0000-0002-0355-0134]{Jessica K. Werk}
\affiliation{Department of Astronomy, University of Washington, 
Seattle, WA 98195, USA}

\author[0000-0003-0789-9939]{Kirill Tchernyshyov}
\affiliation{Department of Astronomy, University of Washington, 
Seattle, WA 98195, USA}

\author[0000-0002-7738-6875]{J. Xavier Prochaska}
\affiliation{Department of Astronomy and Astrophysics, University of California, Santa Cruz, CA 95064, USA}
\affiliation{Kavli Institute for the Physics and Mathematics of the Universe (Kavli IPMU), 5-1-5 Kashiwanoha, Kashiwa, 277-8583, Japan}
\affiliation{Division of Science, National Astronomical Observatory of Japan, 2-21-1 Osawa, Mitaka, Tokyo 181-8588, Japan}

\author[0000-0003-4158-5116]{Yong Zheng}
\affiliation{Department of Physics, Applied Physics and Astronomy, Rensselaer Polytechnic Institute, Troy, NY 12180}

\author[0000-0002-1129-1873]{Mary E. Putman}
\affiliation{Department of Astronomy, Columbia University, New York, NY 10027, USA}

\author[0000-0003-3806-8548]{Drummond B. Fielding}
\affiliation{Department of Astronomy, Cornell University, Ithaca, NY 14853, USA}
\affiliation{Center for Computational Astrophysics, Flatiron Institute, 162 Fifth Avenue, New York, NY 10010, USA}

\author[0000-0002-1468-9668]{Jay Strader}
\affiliation{Center for Data Intensive and Time Domain Astronomy, Department of Physics and Astronomy, Michigan State University, East Lansing, MI 48824, USA}

\begin{abstract}

Metals in the diffuse, ionized gas at the boundary between the Milky Way's interstellar medium (ISM) and circumgalactic medium (CGM), known as the disk-halo interface (DHI), are valuable tracers of the feedback processes that drive the Galactic fountain. 
However, metallicity measurements in this region are challenging due to obscuration by the Milky Way ISM and uncertain ionization corrections that affect the total hydrogen column density. 
In this work, we constrain the ionization corrections to neutral hydrogen column densities using precisely measured electron column densities from the dispersion measure of pulsars that lie in the same globular clusters as UV-bright targets with high-resolution absorption spectroscopy. 
We address the blending of absorption lines with the ISM by jointly fitting Voigt profiles to all absorption components. 
We present our metallicity estimates for the DHI  of the Milky Way based on detailed photoionization modeling to the absorption from ionized metal lines and ionization-corrected total hydrogen columns. 
Generally, the gas clouds show a large scatter in metallicity, ranging between $0.04-3.2\ Z_{\odot}$, implying that the DHI consists of a mixture of gaseous structures having multiple origins.  
We estimate the inflow and outflow timescales of the DHI ionized clouds to be $6 - 35$ Myr. 
We report the detection of an infalling cloud with super-solar metallicity that suggests a Galactic fountain mechanism, whereas at least one low-metallicity outflowing cloud ($ Z < 0.1\ Z_{\odot}$) poses a challenge for Galactic fountain and feedback models.

\end{abstract}

%% Keywords should appear after the \end{abstract} command. 
%% The AAS Journals now uses Unified Astronomy Thesaurus concepts:
%% https://astrothesaurus.org
%% You will be asked to selected these concepts during the submission process
%% but this old "keyword" functionality is maintained in case authors want
%% to include these concepts in their preprints.

%\listoftodos
% {\hypersetup{linkcolor=black}\listoftodos}
%% From the front matter, we move on to the body of the paper.
%% Sections are demarcated by \section and \subsection, respectively.
%% Observe the use of the LaTeX \label
%% command after the \subsection to give a symbolic KEY to the
%% subsection for cross-referencing in a \ref command.
%% You can use LaTeX's \ref and \label commands to keep track of
%% cross-references to sections, equations, tables, and figures.
%% That way, if you change the order of any elements, LaTeX will
%% automatically renumber them.
%%
%% We recommend that authors also use the natbib \citep
%% and \citet commands to identify citations.  The citations are
%% tied to the reference list via symbolic KEYs. The KEY corresponds
%% to the KEY in the \bibitem in the reference list below. 

\section{Introduction} \label{sec:intro}

Gas depletion times of only a few hundred Myrs for star-forming galaxies at $z = 1-3$  have raised the question of how galaxies fuel and sustain their star formation over Gyrs of cosmic evolution \citep{saintonge13}.  The answer is likely to lie in the circumgalactic medium (CGM), a massive reservoir of baryons and metals that extends hundreds of kpc from a galactic disk \citep{tumlinon17}. 
The idea that gas accretion from the CGM continuously fuels the star formation observed in the interstellar medium (ISM) of galaxies has been indirectly supported 
by numerous observations indicating that cool, ionized gas ($T \sim 10^4\ \rm K$) is ubiquitous in low-redshift, L* galaxies out to $\sim$150 kpc with a characteristic total mass of  $M_{\rm cool, CGM} \sim 10^{10-11}\ M_{\odot}$ \citep[e.g.][]{werk14, prochaska17a}. 
At the same time, the CGM  contains metals ejected through outflows by stellar and/or active galactic nuclei (AGN) feedback \citep[e.g.][]{peeples14}, and thus plays an essential role in the baryon cycle that drives galaxy evolution.

The Milky Way is an ideal place to test our understanding of gas accretion, recycling, and outflows (collectively called 
the ``baryon cycle") with numerous, high-quality observational datasets of its multiphase ISM, CGM, and the boundary region between the two, known as the disk-halo interface (DHI). 
Neutral gas clouds moving at radial velocities in the local standard of rest frame (LSR) inconsistent with Galactic rotation have been discovered and catalogued in \ion{H}{1} 21~cm emission surveys \citep[e.g.][]{putman02, kalberla05, winkel16}. 
They are commonly referred to as intermediate-velocity clouds (IVCs; $20 \lesssim |v_{\rm LSR}| \lesssim 90\ \rm km\ s^{-1}$) or high-velocity clouds (HVCs; $|v_{\rm LSR}| > 90\ \rm km\ s^{-1}$) 
and mostly lie at $d \lesssim 20$~kpc \citep{wakker01, putman12, richter17}, placing them in this DHI region.

The infalling cold gas clouds provide valuable information about the gas accretion into the galaxy \citep{lehner11, fox19}. 
Their origin could be outside of the Milky Way such as gas accretion from the intergalactic medium (IGM) \citep{rees77, fumagalli11} or from satellites. 
Alternatively, these cold clouds can form in a ``Galactic fountain" mechanism in which metal-rich outflowing gas driven by feedback processes mixes with metal-poor halo gas, stimulating the halo gas to cool and fall back to the disk \citep{shapiro76, fraternali17}.

In order to understand the origin of gas accretion and the physical processes that power the Galactic fountain, 
we must simultaneously study the ionized and neutral gas at the DHI that exhibit similar line-of-sight velocities. 
Observations of H$\alpha$ emission from warm gas ($T \sim 10^{4-5}\ \rm K$) have revealed its prevalence in the DHI region (d $\lesssim$ 3 kpc) and find that the ionized gas is likely related to the Galactic star formation activities \citep{wang93, haffner03, fang06}. 
The most sensitive probe of this ionized gas at the DHI is ultraviolet (UV) absorption line spectroscopy using background sources like quasars or halo stars. 
UV absorption studies have uncovered that ionized gas spreads out over a larger spatial extent than its kinematically-associated neutral gas \citep[e.g.][]{sembach03, shull09, lehner22}. The ionized gas shows both inflows and outflows, implying that it contains a significant amount of material that could eventually accrete onto the disk and fuel future Galactic star formation \citep{lehner11, lehner12, fox19}.

The detailed kinematic study of ionized gas clouds can provide crucial information about 
their origin and allow us to draw a picture of physical processes at the DHI. 
For example, \cite{fox03} detected several highly ionized gas clouds along a halo O-star line of sight in \ion{O}{6}, \ion{N}{5}, \ion{C}{4}, and \ion{Si}{4}, all having complex absorption-line profiles that vary by ion species. 
While narrow \ion{C}{4} and \ion{Si}{4} lines trace photoionized gas, broader line profiles are indicative of collisionally ionized gas at the boundary between neutral/warm gas clouds and the surrounding hot medium. 
Furthermore, recent kinematic studies of warm, ionized gas at the DHI show that its distribution and kinematics, such as the observed velocity gradients of inflows, can be favorably explained by the Galactic fountain \citep[e.g.][]{bish19, werk19, marasco22}. 
This suggests that the Galactic fountain may play a dominant role not only in driving outflows but also in gas accretion to the Milky Way disk, although further investigation is required.

Another way to constrain the origin of DHI gas is to measure its metallicity. For example, a gas-phase metallicity of a few percent solar would indicate accretion from a more metal-poor source like the IGM, while metallicities closer to solar would imply a  Galactic origin. However, measuring the metallicity of ionized gas at the DHI is challenging due to poor constraints on the total amount of hydrogen in this region. 
Studies that measure the metallicity of neutral HVCs and IVCs have been exclusively done by using \ion{H}{1} 21~cm observations for the clouds with \nhi~$\gtrsim 7 \times 10^{17}\ \rm cm^{-2}$ \citep{wakker99, wakker01, sembach01, collins07}, assuming a negligible ionization correction. 
Yet the ubiquitous ionized gas clouds in this region are dominated by ionized hydrogen rather than neutral hydrogen, preferably having lower \nhi~$\lesssim 10^{17}\ \rm cm^{-2}$ and unlikely to be detected with 
\ion{H}{1} 21~cm observation. Alternatively, several metallicity studies for ionized gas at the DHI 
have been taken with hydrogen Lyman series absorption lines in far-UV (FUV) \citep{fox05, zech08, cashman23}. Given that the ionized hydrogen likely dominates the total hydrogen column, however, there exists
considerable uncertainty in the gas ionization fractions without an independent constraint on ionized hydrogen column density \nhii.

\cite{howk06} were the first to suggest the idea of using the radio dispersion measure (DM) of pulsars to address this issue.  
The pulsar DM provides a precise value of the electron column density $N_e$ which can effectively serve as an independent constraint on
\nhii~with a correction for helium content.  
For sight lines toward globular clusters (GC) that include both UV bright stars and radio pulsars, 
the metallicity of gas can be measured with great precision by combining constraints from UV absorption lines of metal ions, hydrogen Lyman series lines, and the pulsar DM. 
 \cite{howk06} successfully measured the overall metallicity of multiphase gas in the sightline toward M3 with this method. However,  this work did not separately analyze  different gas components at distinct radial velocities (e.g., the Milky Way ISM, IVCs at the DHI) that might have different origins and possibly exhibit variations in metallicity.

In this paper, we build on the method first suggested by \cite{howk06} in order to investigate the metallicity of ionized gas at the DHI for a larger sample of sight lines toward the halo GCs. 
We use archival FUV spectra of UV bright stars in the GCs observed with the Cosmic Origin Spectrograph (COS) 
or the Space Telescope Imaging Spectrograph (STIS) installed on the \textit{Hubble Space Telescope} (\textit{HST}). We describe our data in Section~\ref{sec:data}. 
We separately measure the metallicity of each kinematically-distinct gas cloud by performing Voigt profile fitting (Section~\ref{sec:measurement}) and photoionization modeling (Section~\ref{sec:photoion}) with an independent \nhii~constraint given by the pulsar DM. 
Our results are given in Section~\ref{sec:results}, and we discuss the origin of ionized gas clouds at the Milky Way DHI in Section~\ref{sec:discussion}. 

%-----------------------------------------------------------
\section{Data} \label{sec:data}

%%-- Figure: Spatial Distribution ----------------------
\begin{figure}
  \centering
  \subfloat[XZ plane]{\centering
    \includegraphics[width=\columnwidth]{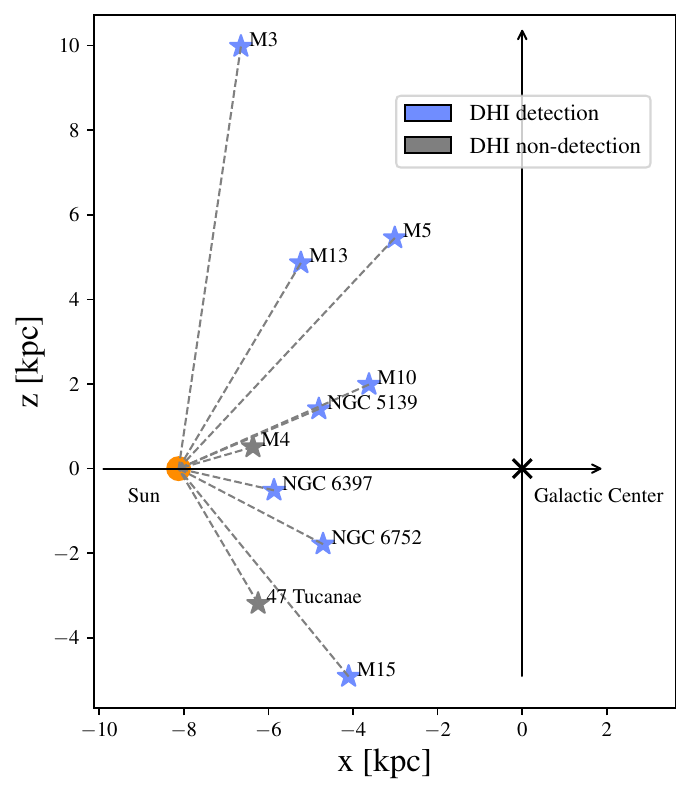}
    \label{fig:xz}}
  \quad
  \subfloat[XY plane]{\centering
    \includegraphics[width=\columnwidth] {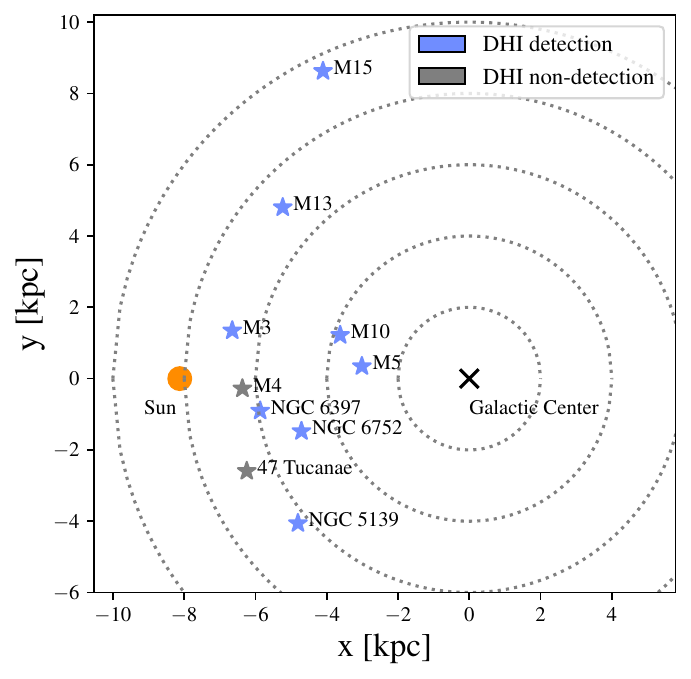}
    \label{fig:xy}}
  \caption{3D distribution of the targeted GCs in the Galactocentric system where the Galactic center lies at (0, 0, 0) kpc and the Sun is at (-8.122, 0, 0) kpc. We mark the GC sight lines where IVCs or/and HVCs are detected in the UV spectrum as ``DHI detection" (blue star).  A ``DHI non-detection" (gray star) indicates that gas kinematically distinct from the Milky Way ISM was not present in the spectrum. }
  \label{fig:target_spatial}
\end{figure}
\begin{deluxetable*}{llccccccccc}
%% This is the title of the table.
\tablecaption{Archival FUV spectral data \label{tab:target}}
\tablehead{\colhead{GC} & \colhead{Star} & \colhead{$l$} & \colhead{$b$} & \colhead{$z_{\rm GC}$} & \colhead{$z_{*}$} & \colhead{Instrument} & \colhead{Grating} & \colhead{S/N} & \colhead{Detection} & \colhead{Proposal} \\ 
\colhead{} & \colhead{} & \colhead{($^{\circ}$)} & \colhead{($^{\circ}$)} & \colhead{(kpc)} & \colhead{(kpc)} & \colhead{} & \colhead{} & \colhead{} & \colhead{} & \colhead{ID} }

\rotate
%% All data must appear between the \startdata and \enddata commands
\startdata
M5 (NGC 5904)       &  NGC5904-KUST723 &  3.86 &  46.79 &  5.47 &  $4.12_{-1.36}^{+3.84}$ &STIS &  E140M &  29.81 & \cmark & 9410 \\
  & \dotfill & \dotfill & \dotfill & \dotfill & \dotfill &  FUSE &  SiC2A/LiF1A/LiF2A &  10.94 & \dotfill & D157 \\
M10 (NGC 6254)      &  NGC6254-ZNG1 &  15.16 &  23.09 &  1.74 & $2.68_{-0.32}^{+0.42}$ &  COS &  G130/160M &  41.48 & \cmark & 13721 \\
M3 (NGC 5272)      &  NGC5272-ZNG1 &  42.50 &  78.68 &  10.02 & $10.94_{-2.55}^{+3.37}$ & COS &  G130/160M &  36.98 &  \cmark & 11527 \\
  & \dotfill & \dotfill & \dotfill & \dotfill & \dotfill &  FUSE &  SiC2A/LiF1A/LiF2A &  13.96 & \dotfill & P101 \\
M13 (NGC 6205)     &  NGC6205-ZNG1 &  58.97 &  40.94 &  4.67 & $5.90_{-1.27}^{+1.76}$ & COS &  G130/160M &  56.60 &  \cmark & 11527 \\
  & \dotfill & \dotfill & \dotfill & \dotfill & \dotfill & FUSE &  SiC2A/LiF1A/LiF2A &  12.02 & \dotfill & P101 \\
M15 (NGC 7078)     &  NGC7078-ZNG1 &  65.04 &  -27.29 &  -4.76 & \dotfill & COS &  G130/160M &  40.90 &  \cmark & 11527 \\
  & [J76] B212734.4+115714 & 65.02 & -27.31 & -4.76 & $-2.32_{-1.24}^{+0.71}$ & FUSE &  SiC2A/LiF1A/LiF2A &  8.50 & \dotfill & D157 \\
47 Tucanae (NGC 104) &  NGC104-BS &  305.91 &  -44.88 &  -3.24 & $-3.65_{-2.18}^{+1.28}$ & COS &  G130/160M &  36.63 & \xmark & 13721 \\
$\omega$ Centauri (NGC 5139) &  NGC5139-ROA-5701 &  309.24 &  15.05 &  1.36 &  $1.62_{-0.20}^{+0.33}$ & COS &  G130/160M &  40.32 &  \cmark & 12032 \\
NGC 6752 &  NGC6752-BUON-1754 &  336.62 &  -25.64 &  -1.72 & $-2.92_{-0.80}^{+0.72}$ & COS &  G130/160M &  57.56 &  \cmark & 12032 \\
  & \dotfill & \dotfill & \dotfill & \dotfill &  \dotfill & FUSE &  SiC2A/LiF1A/LiF2A &  7.46 & \dotfill & C076 \\
NGC 6397 &  NGC6397-ROB162 &  338.19 &  -11.94 &  -0.46 & $-0.48_{-0.03}^{+0.03}$ &  STIS &  E140M &  41.94 &  \cmark & 9410 \\
  & \dotfill & \dotfill & \dotfill & \dotfill &  \dotfill & FUSE &  SiC2A/LiF1A/LiF2A &  7.22 & \dotfill & A026 \\
M4 (NGC 6121)      &  NGC6121-Y453 &  350.98 &  16.05 &  0.62 & $0.58_{-0.06}^{+0.06}$ & COS &  G130/160M &  32.53 & \xmark & 13721 \\
\enddata
\tablecomments{The 10 sight lines toward the halo GCs of which both pulsar DM and 
UV spectral data are available. The Galactic coordinates ($l, b$) and height from the disk of the targeted GCs ($z_{\rm GC}$) and UV bright stars ($z_{*}$) are presented. We detect IVCs and/or HVCs for 8 of the sightlines (checkmark). All the data presented here can be found in MAST:\dataset[https://doi.org/10.17909/k15n-zp34]{https://doi.org/10.17909/k15n-zp34}.}
\end{deluxetable*}
%%-----------------------------------
\subsection{Target Selection}
To select targets for this experiment, we first referred to a comprehensive pulsar catalog \footnote{\url{https://www3.mpifr-bonn.mpg.de/staff/pfreire/GCpsr.html}} by Paulo Freire to obtain a list of GCs for which pulsar DMs are available. 
The pulsar DM can be converted to the electron column, and serves as a strict upper limit on the total electron column density of ionized gas along the sight line. Ultimately, this electron column will be converted to a constraint on the \ion{H}{2} column density of ionized gas at the Milky Way DHI that will in turn, allow us to measure the metallicity at the DHI. 
The electron column densities for each line of sight are presented in Table~\ref{tab:appendix_measure} in Appendix~\ref{sec:appendix} with references.

We performed an archival search in the Barbara A. Mikulski Archive for Space Telescopes (MAST) for existing high-resolution FUV spectra of UV bright stars in those GCs. Table~\ref{tab:target} lists the ten lines of sight toward globular clusters with pulsar DM measurements for which we were able to obtain FUV spectra. In order to ensure the cluster membership of the UV bright stars, we only selected the targets of which membership is well studied \citep[e.g.][]{chayer15, moehler19, dixon19}, and confirmed that their (photo)geometric distances estimated with \textit{Gaia} DR3 \citep{bailer-jones21} are comparable to the distances to the GCs \citep{baumgardt21}.

The selected targets all lie at Galactic heights of $z \lesssim 10$ kpc (Table~\ref{tab:target}) and thus serve as lines of sight sensitive to the DHI. 
Their spatial distribution with respect to the Milky Way disk is shown in Fig.~\ref{fig:target_spatial}. We note that, here, we refer to and use the distance to the clusters as the distance to the targets, considering the confidence in their cluster membership and errors on the order of a few kpc in the Gaia distances to the individual stars.

%----------------------------------
\subsection{\textit{HST}/COS and STIS}
The majority of our archival data are FUV spectra observed with the \textit{HST}/COS G130M and 160M gratings. The wavelength coverage spans 1132-1775~\AA\ with a characteristic spectral resolution of $R \sim 12,000 - 20,000$. The relatively high S/N (S/N $>$ 30) of the FUV spectra allows us to analyze a number of absorption lines from low ions (e.g., \ion{O}{1}, \ion{C}{2}, \ion{Si}{2}, \ion{Si}{3}) and intermediate ions (e.g., \ion{Si}{4}, 
\ion{C}{4}, \ion{N}{5}) that trace the warm ionized gas at the DHI. 

For the two sight lines M5 and NGC~6397, \textit{HST}/STIS E140M data are available.  These high S/N spectra cover wavelengths of 1144–1730~\AA\ with a characteristic spectral resolution of $R \sim 30,000$, corresponding to a velocity resolution of $\sim 10\ \rm km\ s^{-1}$. We continuum-normalize all COS co-added spectra from the HSLA \citep{HSLA} and STIS spectra using ``lt\_continuumfit" a Python tool in the open-source package \texttt{Linetools} \citep{prochaska16}.

\subsection{FUSE}
\textit{Far Ultraviolet Spectroscopic Explorer (FUSE)} spectra are available for six lines of sight (Table~{\ref{tab:target}}). 
The spectra were obtained with the SiC2A, LiF1A, and LiF2A channels, with segments that cover 916-1005~\AA, 987-1082~\AA, and 1086-1181~\AA,  respectively, with a spectral resolution of $R \sim 10,000 - 20,000$. 
This wavelength range covers additional metal ions such as \ion{S}{3} and \ion{Fe}{3} that are unavailable in the COS or STIS spectra.

Furthermore, \textit{FUSE} spectra provide a constraint on \ion{H}{1} column density. Although we focus on ionized clouds close to the disk where \ion{H}{2} likely accounts for a significant fraction of the total hydrogen, the \ion{H}{1} column density (\nhi) is an important, additional constraint on the photoionization modeling that is discussed in Section~\ref{sec:photoion}.

%% COS diameter: 2.5"
%% STIS diameter: 0.2" (0.2 x 0.2)
%% EBHIS ~ 10.8

A way to precisely to measure \ion{H}{1} column density of the absorbing clouds is using \ion{H}{1} Lyman series absorption lines in high-resolution FUV spectra. 
Although recent all-sky \ion{H}{1} 21~cm surveys such as the Leiden/Argentine/Bonn survey 
(LAB; \citealt{kalberla05}) and the HI4PI survey \citep{hi4pi16} have been successfully used to measure \nhi~of neutral-gas bearing clouds in emission, these data offer only coarse constraints on the \nhi~ for two reasons. 
First of all, these surveys have an effective beam size of 10 - 30\arcmin~ which is at least several tens of times larger than the HST/COS aperture (2.5\arcsec) or STIS (0.2\arcsec) and possibly gives a larger or smaller value of \nhi~due to beam smearing \citep{howk06}. 
Additionally, we were not able to detect any possible IVCs and HVCs from both LAB and EBHIS data for most of our sample lines of sight, indicating that the ionized clouds we are finding either have \nhi~below the detection limit of \ion{H}{1} surveys of \nhi~$\gtrsim 10^{18}\ \rm cm^{-2}$ and/or are small and thus spatially-unresolved \citep{richter17}. 
We finally note that \nhi~measurements on the order of $10^{15-18}\ \rm cm^{-2}$ would be consistent with several previous studies on ionized gas clouds at the DHI  \citep{collins04, zech08, cashman23}. Therefore, we use the FUSE spectra to measure \ion{H}{1} Lyman series absorption lines with wavelengths shorter than Ly$\gamma$ down to Ly$\mu$ at 916~\AA\ to probe \nhi~of $\sim 10^{14-16}\ \rm cm^{-2}$. 
%------------------------------------

%------------------------
\section{Ion Column Density Measurements} \label{sec:measurement}

%%------------------------------------
\begin{figure*}
    \centering
    \includegraphics[width=\textwidth]{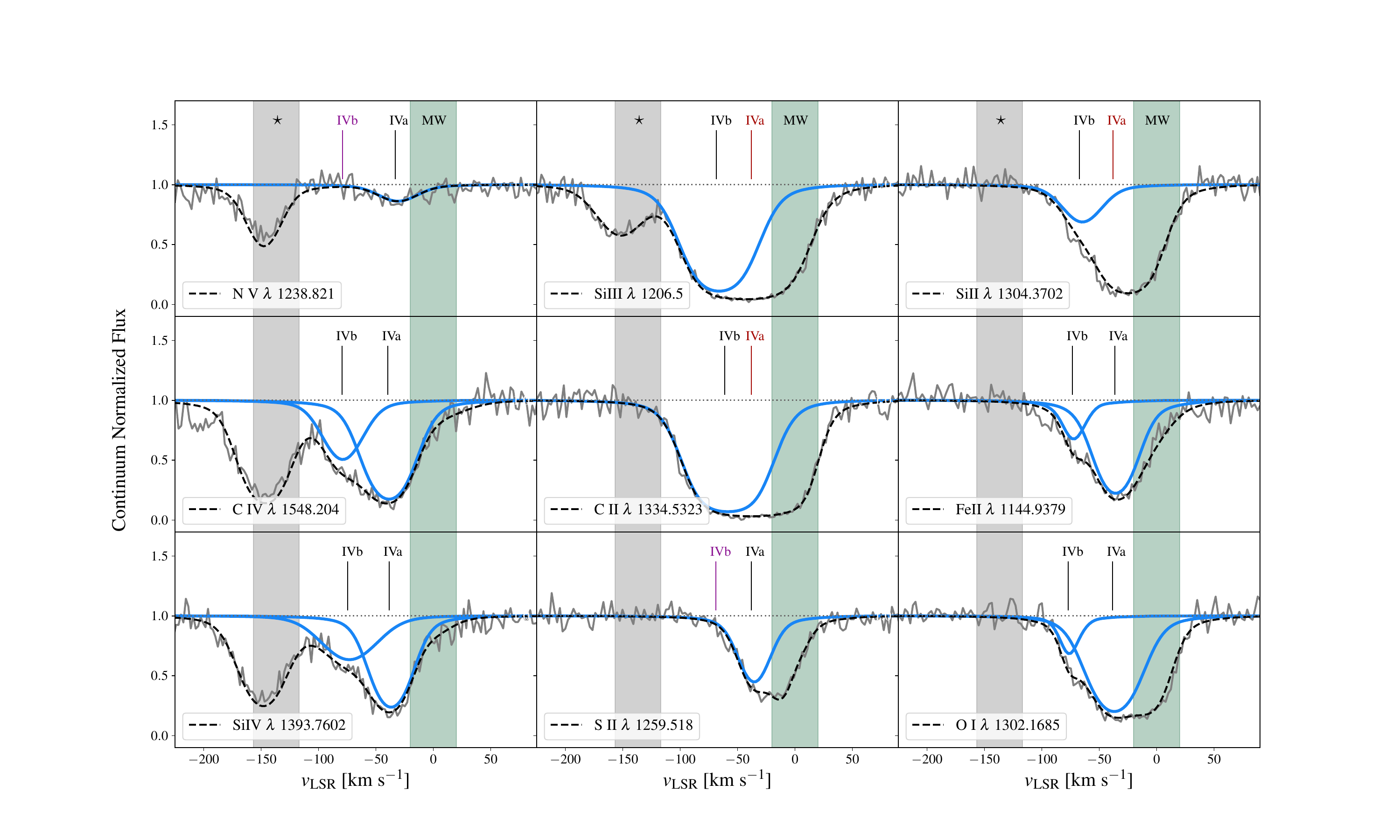}
    \caption{The Voigt profile fit for the M3 sight line (NGC5272-ZNG1). 
    The Black dashed line shows the joint fit including all the absorption components. 
    We separate two intermediate velocity gas clouds, IVa and IVb (blue-solid lines), from the Milky Way ISM (green-shaded) and star (gray-shaded). The component labeled IVa is well-detected in highly-ionized metal species, but is
    severely blended with saturated absorption from the Milky Way in a few low-ionization metal species (red labels), which makes it impossible to constrain gas column densities from those absorption lines. We mark IVC non-detections in a given ion species with a purple label.}
    \label{fig:vpf}
\end{figure*}

We apply Voigt profile fitting to the ion absorption lines detected in the continuum-normalized UV spectra to precisely measure the ion column densities. 
In order to measure the metallicity of ionized gas clouds at the DHI, 
the absorption features that originate from ionized gas clouds need to be separated from the Milky Way ISM absorption or stellar absorption. 
Because the only distance information on gas clouds is an upper limit constraint from the background GCs, we identify the absorption features in velocity space, in the LSR rest frame. 
The components with $|v_{\rm LSR}| \leq 20~{\rm km\ s^{-1}}$ are classified as the Milky Way ISM absorption, and the components with $|v_{\rm LSR} - v_{\rm star}| \leq 20~{\rm km\ s^{-1}}$ are regarded as stellar absorption. 
We adopt the terminology of IVC and HVC to refer to components of which 
absolute LSR velocity is $20 < |v_{\rm LSR}| \leq 90~{\rm km\ s^{-1}}$ and 
$|v_{\rm LSR}| > 90~{\rm km\ s^{-1}}$, respectively \citep{wakker01, putman12, werk19}.
For defining IVCs, a lower velocity limit of $40\ \rm km\ s^{-1}$ is more generally accepted, as velocities below this threshold can be indistinguishable from galactic rotation on certain sight lines \citep[e.g.][]{vanwoerden04}. However, we adopt a lower value of $20\ \rm km\ s^{-1}$ to maximize the detection of clouds distinguishable from the ISM and to maintain a consistent threshold across all samples.

We identify all the absorption components within $|v_{\rm LSR}| < 400~{\rm km\ s^{-1}}$ using \texttt{pyigm\footnote{\url{https://github.com/pyigm/pyigm}} IGMGuesses} GUI, 
which shows each absorption line in velocity space, allowing us to easily identify components by eye using velocity offsets. 
Also, it makes the first estimate of Voigt profile parameters for each ion species: ion column density $N_{\rm ion}$, Doppler parameter $b$, and $v_{\rm LSR}$. 
Some absorption features are saturated; given that the line-spread function of the COS instrument features a broad core and extended wings, compared to Gaussian, the line profile convoluted with the COS line-spread function typically exhibits a peak value about 0.8 times that of an equivalent Gaussian profile \citep{ghavamian09}. Therefore, we defined a feature as saturated if there were three or more consecutive pixels with flux values below the normalized flux of 0.2. 
If multiple lines are available for the ion species, 
we select the line with the lowest oscillation strength for the identification to avoid saturation and minimize uncertainty in the measurements.

%------------------------------
\begin{deluxetable}{lcccc}
   % \tablewidth{\textwidth}
   \tabletypesize{\footnotesize}
    \tablecaption{Voigt profile measurements and ion column densities for the M3 line of sight. \label{tab:measurements}}
    \tablehead{\colhead{Comp.} & \colhead{Ion} & \colhead{$\log{N}$} & \colhead{$b$} & \colhead{$v_{\rm LSR}$} \\ 
    \colhead{} & \colhead{} & \colhead{($\rm cm^{-2}$)} & \colhead{($\rm km\ s^{-1}$)} & \colhead{($\rm km\ s^{-1}$)} }
    \startdata
    Total & $e^{-}$ & $<$ \tablenotemark{a} 19.91 $\pm$ 0.01 & \dotfill & \dotfill \\
    IVa & \ion{H}{1}  & $>$ 16.96 $\pm$ 0.16 & \dotfill & \dotfill \\
    & \ion{O}{1}  & $>$ 14.81 $\pm$ 0.07 & 21.8 $\pm$ 5.3 & -38.3 $\pm$ 2.7\\
    & \ion{Fe}{2} & 14.92 $\pm$ 0.02 & 27.9 $\pm$ 0.8 & -37.5 $\pm$ 0.7\\
    & \ion{S}{2}  & 14.83 $\pm$ 0.04 & 12.5 $\pm$ 1.9 & -38.0 $\pm$ 1.7\\
    & \ion{Fe}{3} & 14.40 $\pm$ 0.08 & 19.8 $\pm$ 2.9 & -45.8 $\pm$ 1.7\\
    & \ion{S}{3} & 14.41 $\pm$ 0.10 & 21.9 $\pm$ 3.5 & -23.7 $\pm$ 4.1\\
    & \ion{Si}{4} & 13.63 $\pm$ 0.11 & 16.9 $\pm$ 3.1 & -38.2 $\pm$ 1.8\\
    & \ion{C}{4}  & 14.21 $\pm$ 0.06 & 19.6 $\pm$ 3.5 & -39.6 $\pm$ 1.7\\
    & \ion{N}{5}  & 13.15 $\pm$ 0.05 & 19.1 $\pm$ 4.3 & -33.2 $\pm$ 2.6\\
    IVb & \ion{H}{1}  & $>$ 16.92 $\pm$ 0.16 & \dotfill & \dotfill \\
    & \ion{O}{1}  & 13.92 $\pm$ 0.29 & 8.6 $\pm$ 5.7 & -73.1 $\pm$ 4.1\\
    & \ion{Fe}{2} & 14.02 $\pm$ 0.12 & 4.0 $\pm$ 0.9 & -73.3 $\pm$ 1.0\\
    & \ion{Si}{2} & 14.23 $\pm$ 0.06 & 26.2 $\pm$ 2.7 & -76.7 $\pm$ 3.2\\
    & \ion{S}{2}  & $<$ 14.63 $\pm$ 0.03 & \dotfill & \dotfill \\
    & \ion{C}{2}  & $>$ 14.64 $\pm$ 0.22 & 27.0 $\pm$ 3.4 & -61.9 $\pm$ 7.5\\
    & \ion{Fe}{3} & 13.82 $\pm$ 0.23 & 21.9 $\pm$ 7.9 & -81.4 $\pm$ 8.9\\
    & \ion{Si}{3} & $>$ 13.61 $\pm$ 0.28 & 25.2 $\pm$ 5.3 & -68.4 $\pm$ 11.0\\
    & \ion{S}{3} & 13.75 $\pm$ 0.45 & 18.0 $\pm$ 9.1 & -56.3 $\pm$ 11.5\\
    & \ion{Si}{4} & 13.17 $\pm$ 0.30 & 28.0 $\pm$ 13.0 & -74.6 $\pm$ 15.6\\
    & \ion{C}{4}  & 13.64 $\pm$ 0.14 & 17.8 $\pm$ 4.9 & -79.7 $\pm$ 4.9\\
    & \ion{N}{5}  & $<$ 12.55 $\pm$ 0.19 & \dotfill & \dotfill 
    \enddata
    \tablenotetext{a}{Pulsar DM by \cite{hessels07}}
    \tablecomments{Properties of all the detected absorption lines in the sight line toward M3. The two IVCs are detected at $v_{\rm LSR} \sim -37\ \rm km\ s^{-1}$ and $\sim -70\ \rm km\ s^{-1}$ which are IVa and IVb, respectively. The electron column density given by pulsar DM is the total column density for this sight line. $\log{N}$ is ion column density, and $b$ is the Doppler parameter of the absorption line.}
\end{deluxetable}
%---------------------------------------------

We detect HVCs and/or IVCs in the sight lines toward 8 of 10 GCs. 
We consider it a detection when absorption components of at least three different ions are identified for robustness. 
We note that, despite the lower velocity threshold of $|v_{\rm LSR}| < 20\ \rm km\ s^{-1}$, all detected IVCs are found at $|v_{\rm LSR}| > 35\ \rm km\ s^{-1}$, aligning closely with the more commonly considered threshold (Table~\ref{tab:appendix_measure} in Appendix~\ref{sec:appendix}). 
With the initial values given by \texttt{IGMGuesses}, we jointly fit Voigt profiles for every absorption component using a public python 
package \texttt{veeper\footnote{The original version of \texttt{veeper}: \url{https://github.com/jnburchett/veeper}. 
In this work, we use a custom version (\url{https://github.com/mattcwilde/veeper}).}}. 
We present the best-fit Voigt profiles for the M3 sight line (NGC5272-ZNG1) as an example in Fig.~\ref{fig:vpf}. 
In Fig.~\ref{fig:vpf}, the green-shaded band is the Milky Way ISM absorption, and the gray-shaded 
band shows stellar absorption. For this sight line, 
we detect two IVCs, `IVa' at $v_{\rm LSR} \sim -37\ \rm km\ s^{-1}$ and `IVb' at 
$\sim -70\ \rm km\ s^{-1}$. Blue-solid lines show the best-fit Voigt profile for each component marked with black labels. 
We also mark `no-information' absorption lines with the red label in cases of severely blending with the Milky Way ISM and 
`non-detections' with the purple label. Table~\ref{tab:measurements} presents the corresponding 
best-fit parameters of the Voigt profile for each ion species. The ion column density measurements determined by fitting 
Voigt profiles for all the sight lines are presented in Table~\ref{tab:appendix_measure} in Appendix~\ref{sec:appendix}. 
We present lower or upper limits for several ions because of saturation or non-detection, respectively. 
For the saturated lines, the measured column density is taken as a lower limit. 
In the case of a non-detection, an upper limit is estimated by the apparent optical depth method (AODM; \citealt{savage91}) with the $2\sigma$ level of significance (95 \% confidence). 
Please refer to \cite{bowen08} for the detailed method.

We also constrain \nhi~using \ion{H}{1} Lyman series lines detected in absorption for M5, M3, M15, and NGC~6397 when a \textit{FUSE} spectrum is available. 
We measure \ion{H}{1} column density using the AODM applied to the Lyman absorption lines. Similar to the approach taken for other ion species, we select lines with the lowest oscillation strengths, specifically at $\lambda 916, \lambda 917$, and $\lambda 918$, to minimize saturation. Nevertheless, we consider these column density estimates as lower limits because the lines are still found to be saturated, and Voigt profile fitting is not appropriate considering their low S/N (Table~\ref{tab:target}).
% Although we select the least saturated lines at $\lambda 916, \lambda 917$ and $\lambda 918$, we only estimate the lower limits using the AODM because they are still found to be saturated, and Voigt profile fitting is not appropriate considering their low S/N (Table~\ref{tab:target}).

%-----------------------------------------
\section{Photoionization Modeling} \label{sec:photoion}

%--------------------------------
\begin{figure}
    \centering
    \includegraphics[width=0.45\textwidth]{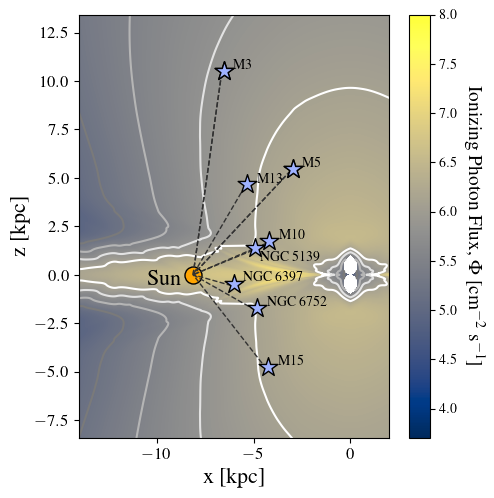}
    \caption{The Milky Way radiation field modeled by \cite{fox05}, with the positions in xz space of our 8 targets marked by blue stars. The color scale indicates the level of the assumed ionizing photon flux, Log $\Phi$, in units of  photons cm$^{-2}$ s$^{-1}$. }
    \label{fig:mw_radiation}
\end{figure}
%--------------------------------

Since the majority of detected ion species in our samples are low ions that may have undergone photoionization, 
we conduct photoionization modeling to correct for ionization effects, characterize the ionized clouds, and ultimately estimate their metallicity. 
We also detect intermediate ions, \ion{Si}{4}, \ion{C}{4}, and \ion{N}{5}, 
that may be part of the photoionized warm gas ($T \sim 10^{4-5}\ \rm K$) or also may be in a collisionally-ionized transitional phase at $\sim 10^5\ \rm K$ \citep{werk19}. 
Photoionization modeling is a valuable tool for understanding the behavior of intermediate ions. 
By applying a photoionization model, we can determine the extent to which the observations of the intermediate ions can be explained by photoionization, and evaluate the need for other mechanisms (e.g. turbulent mixing layers, shock ionization) to account for discrepancies.

The spectral synthesis code CLOUDY \citep{ferland17} is used for photoionization modeling. 
We use the Milky Way radiation model developed by \cite{fox05} as the ionizing source that includes stellar radiation and extragalactic background radiation (Fig.~\ref{fig:mw_radiation}). 
We generate a grid of gas cloud models parameterized by neutral hydrogen column density $N_{\rm HI}$, 
metallicity $Z/Z_{\odot}$, and ionization parameter $U$, 
where the ionization parameter $U$ is defined as the ratio between the number density of ionizing photons and hydrogen ($U \equiv n_{\gamma} / n_{\rm H} = \Phi / c n_{\rm H}$). 
The ionization parameter $U$ is useful for our models because it inherently scales with both the ionizing photon flux and the hydrogen number density. Thus, $U$ effectively incorporates the variations in ionizing flux that may arise from uncertainties in the distances to the clouds. Our grid spans $\log{N_{\rm HI}} = 14.5$ to 20 in steps of 0.25 dex, 
$\log{Z/Z_{\odot}} = -2$ to +1 in steps of 0.1 dex, and  $\log{U} = -4$ to 0 in steps of 0.25 dex. 
The main assumptions made are that gas clouds (1) are in thermal and ionization equilibrium, (2) have a plane-parallel slab geometry with a uniform density, 
and (3) have relative solar abundances of the elements. We discuss the systematic uncertainties inherent in this method in Section \ref{subsec:low ion result}.

\begin{figure}
    \centering
    \includegraphics[width=0.4\textwidth]{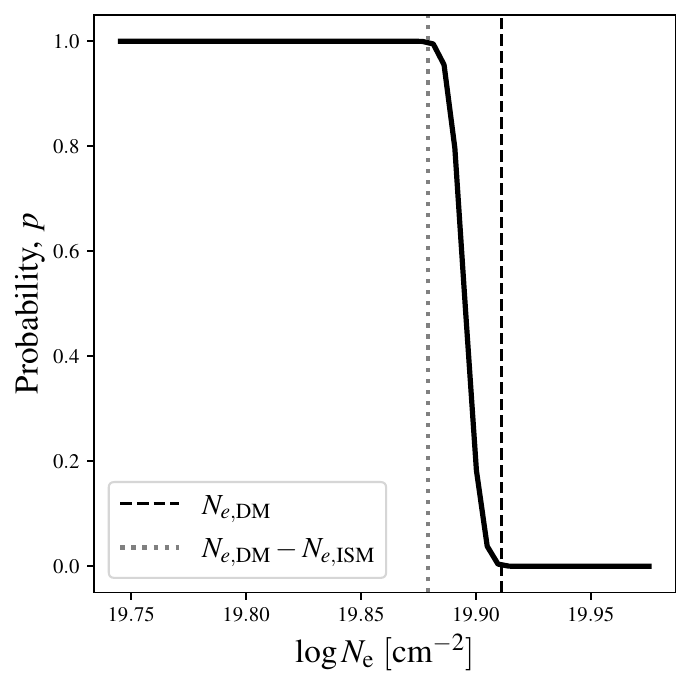}
    \caption{The applied likelihood of $N_{\rm e}$ for the M3 sight line.}
    \label{fig:Ne_cdf}
\end{figure}

We apply a Hamiltonian Markov chain (HMC) approach to estimate metallicity by comparing a linearly interpolated grid model to observational constraints, which are ion column densities determined in Sec~\ref{sec:measurement} including $N_{\rm HI}$. 
We refer to the method described in \cite{fumagalli16} which worked successfully for estimating metallicities of the CGM of low-redshift galaxies \citep{prochaska17b}.

Furthermore, our strategy to find the best-fit photoionization model and estimate metallicity uses the electron column density from pulsar DMs, $N_{e, \rm DM}$, as a strict upper limit of $N_{e}$ of a gas cloud. Given that about 4-8~\% of the ISM is ionized \citep{deavillez12, jenkins13}, we subtract the approximate contribution from the ISM to $N_{e, \rm DM}$ using the NE2001 Galactic free electron density model \citep{cordes02}. 
This model provides an electron column density for the given sight line with a distance. 
Since the model also includes the contribution from the thick disk component that may be a part of the DHI, we exclude the thick disk component from the model and estimate the ISM contribution, $N_{e, \rm ISM}$. 
The model electron column density, $N_{e, \rm model}$, is directly calculated with a helium correction by the CLOUDY model ($N_{e, \rm model} = N_{\rm HII} + N_{\rm HeII} + + 2N_{\rm HeIII}$). 
The likelihood of $N_e$ is defined as Q-function with the given $N_{e, \rm DM}$ and $N_{e, \rm ISM}$ (Fig.~\ref{fig:Ne_cdf}). 
Fig 4 shows the probability function of Ne, and that the subtraction of the ISM based on NE2001 extends the transition region in the Q-function. 
With internally self-consistent ionization corrections and helium corrections, this method effectively constrains the ionized hydrogen column density \nhii, which is expected to dominate in mass over \ion{H}{1} in the diffuse gas around the Milky Way \citep{putman12}.

For the sight lines of M3 and M5, we detect multiple clouds with different velocity offsets. 
Since $N_{e, \rm DM}$ is the upper limit of the total electron column density of all the clouds in the same line of sight, we perform a joint HMC fitting for the gas clouds in order to satisfy the constraint on the total electron column density. 
For example, two IVCs, IVa and IVb, were detected in the M3 line of sight. Thus, the upper limit of the total electron column density $N_{e, \rm tot}$ is given by 
$N_{e, \rm DM} >N_{e, \rm tot} = N_{e, \rm IVa} + N_{e, \rm IVb}$, where $N_{e, \rm IVa}$ and $N_{e, \rm IVb}$ are the model electron column density of IVa and IVb, respectively. 
We validate the self-consistency by ensuring that $N_{e, \rm tot}$ estimated by the best-fit model is below $N_{e, \rm DM}$. 
Finally, in order to determine the ionization parameter, 
add constraints from ion column density measurements with a wide range of ionization potentials, 
specifically including a series of ion species of an element (e.g., \ion{Si}{2}, \ion{Si}{3}, and \ion{Si}{4}, \ion{Fe}{2} and \ion{Fe}{3}).

For each target, we perform 4 HMC runs of 12,000 steps each, discarding the first 3,000 steps, for the full range of the model grid. 
The probabilistic programming library \texttt{NumPyro} is used for the HMC fitting processes \citep{bingham2019pyro, phan2019composable}. 

%----------------------
\begin{figure*}
    \centering
    \includegraphics[width=\textwidth]{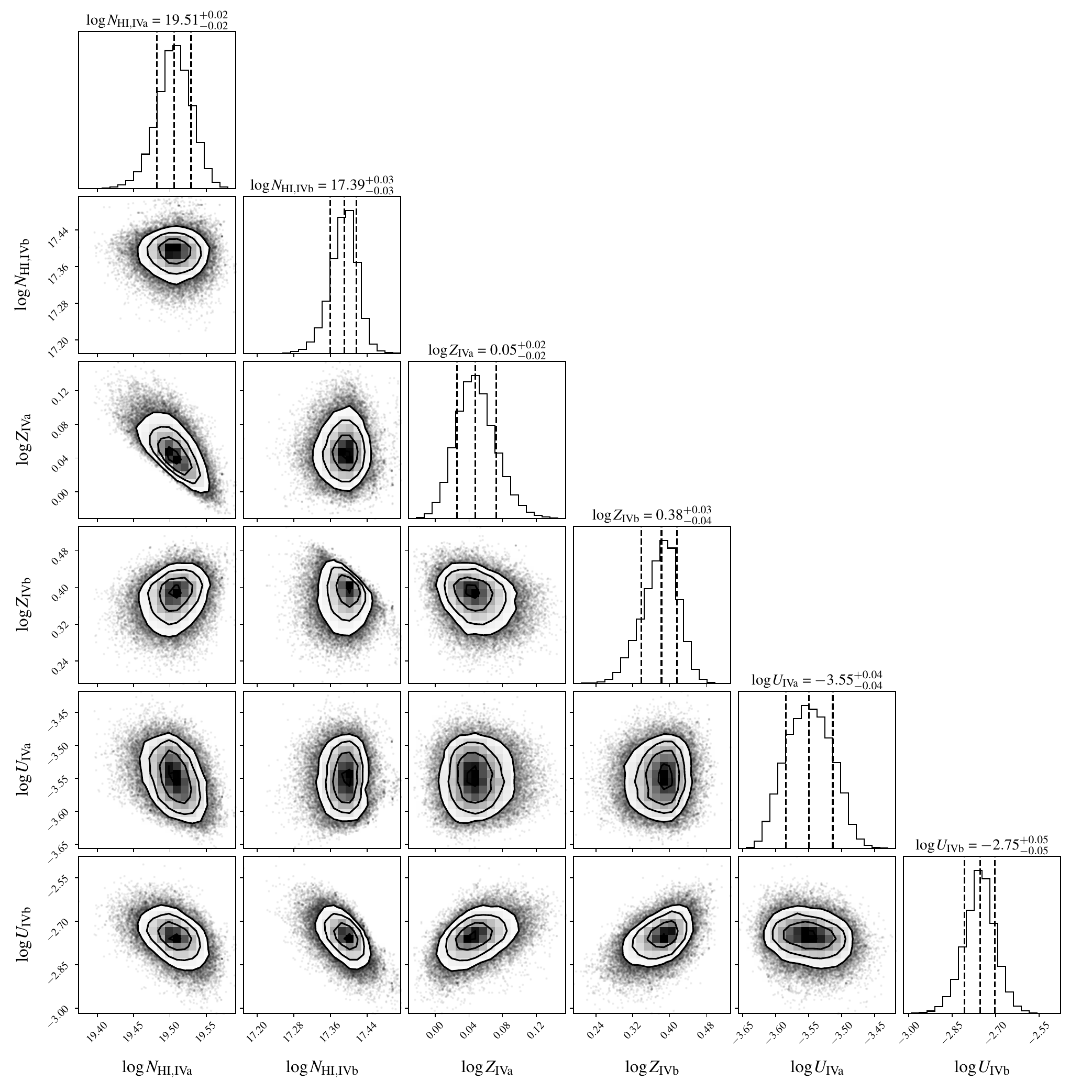}
    \caption{The HMC corner plot of M3 fitting to the photoionized low-ion cloud model. The fitted parameters are neutral hydrogen column density \nhi, metallicity $Z/Z_{\odot}$, and ionization parameter $U$. The median values of posteriors for each IVC are presented with $\pm 1\sigma$.}
    \label{fig:corner}
\end{figure*}
%------------------------

\begin{figure*}
  \centering
  \subfloat[The best-fit photoionization model for the IVa cloud detected along the sight line toward M3]{\centering
    \includegraphics[width=0.65\textwidth]{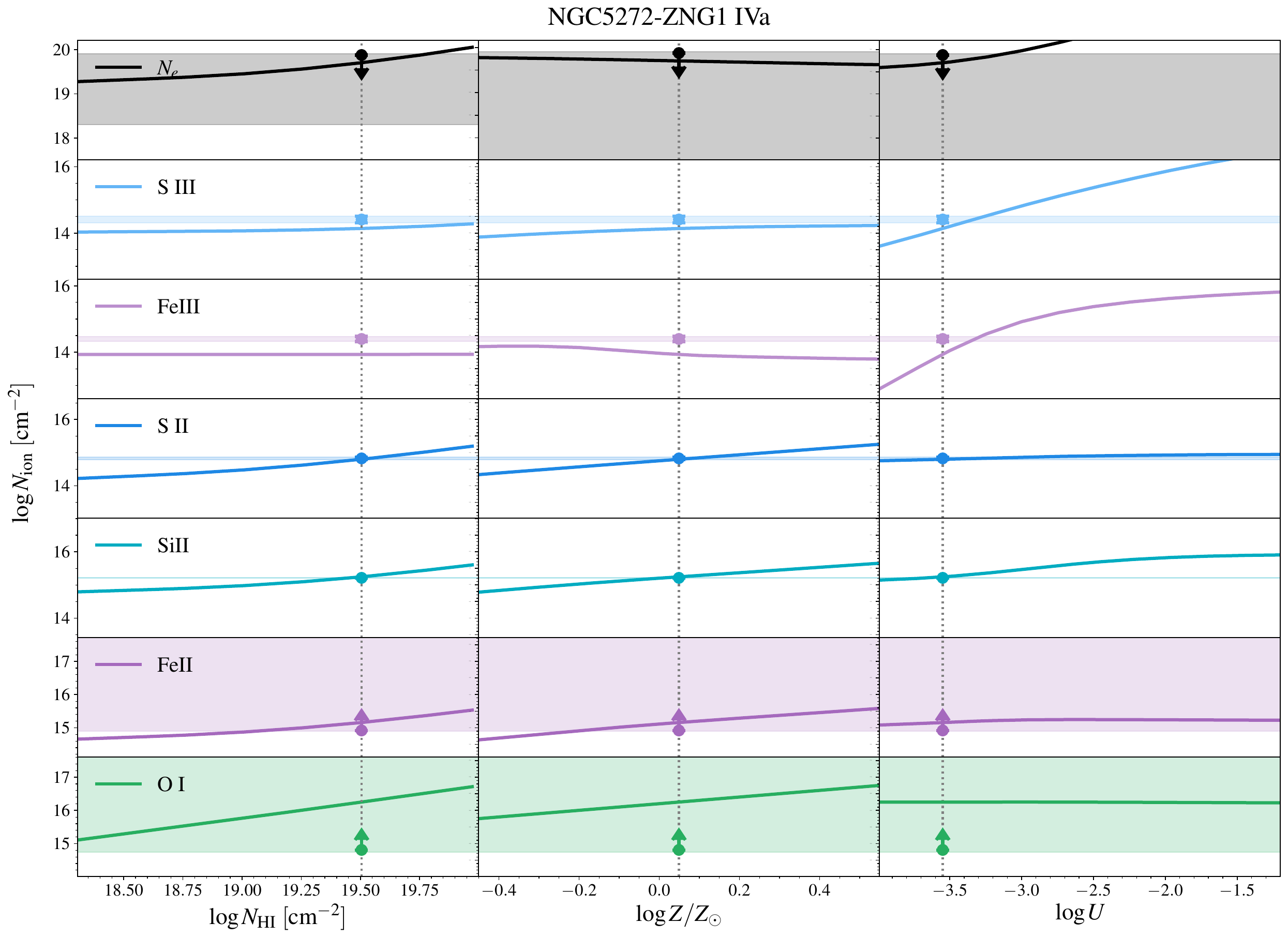}
    \label{fig:fig1}}
  \quad
  \subfloat[The best-fit photoionization model for the IVb cloud detected along the sight line toward M3]{
    \includegraphics[width=0.65\textwidth]{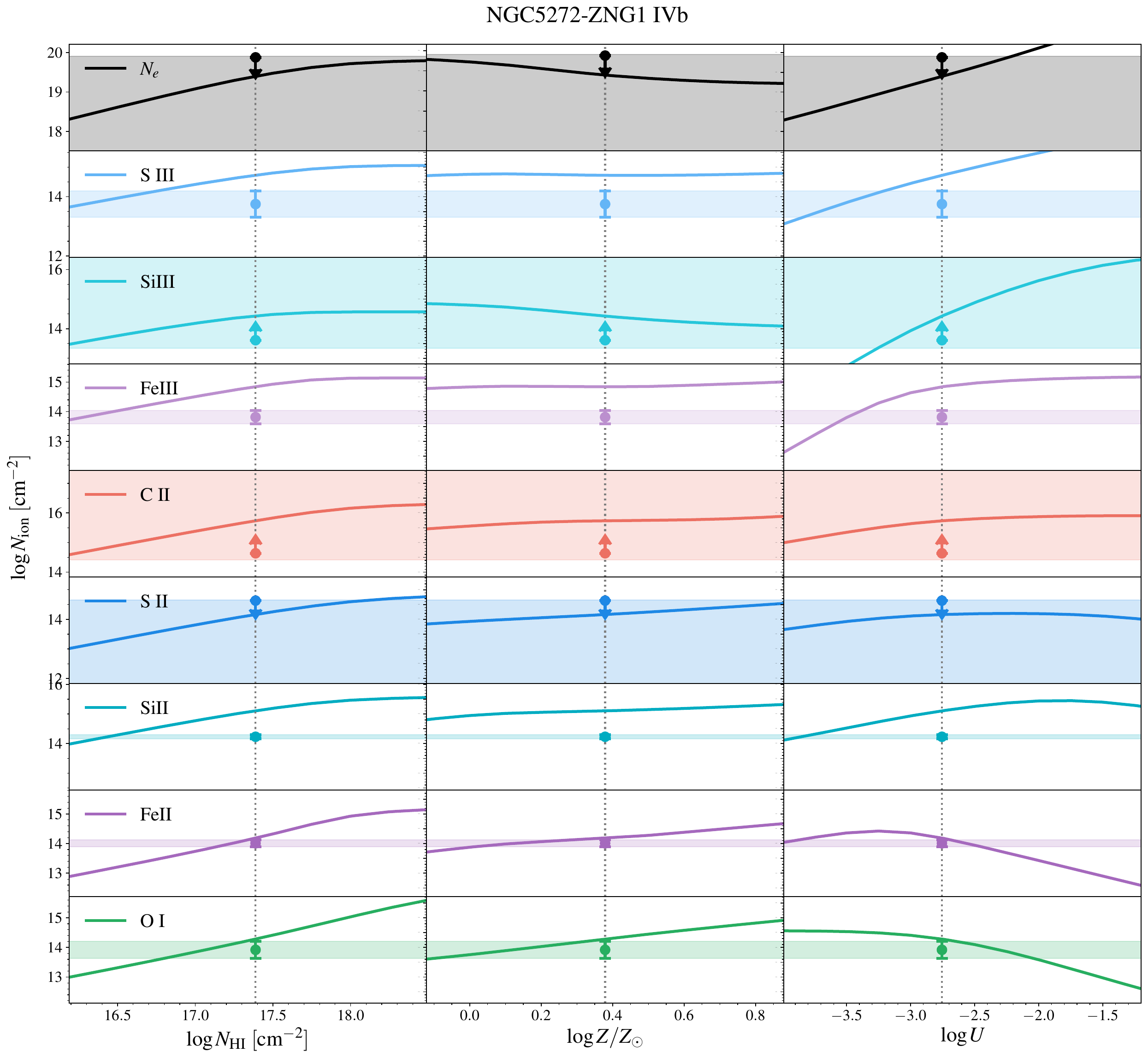}
    \label{fig:fig2}}
  \caption{The best-fit CLOUDY photoionization models for low ion clouds (solid lines) and ion column density $N_{\rm ion}$ measurements (points and shaded regions, showing the errors) for IVa and IVb detected along the M3 sight line. Each column shows how the CLOUDY model varies by a single parameter, with the other two parameters held at the best fit values from the HMC analysis. The model parameters we vary are \ion{H}{1} column density \nhi~(left), metallicity $Z/Z_{\odot}$ (middle), and ionization parameter $U$ (right). The best-fit value for each parameter is presented by the gray-dotted vertical line. We mark upper and lower limits with arrows.}
  \label{fig:spaghetti}
\end{figure*}

We implement HMC fitting against several cloud models to assess the degree to which photoionization can account for the observed data. The considered cloud models are as follows:
\begin{itemize}
    \item Photoionized low-ion cloud (``Low ion''): We assume the detected intermediate ions such as \ion{Si}{4}, \ion{C}{4}, and \ion{N}{5} arise from a transitional-temperature gas phase at $\sim 10^5$~K, and are collisionally ionized. In this case, we apply the photoionization modeling to only the low-ion species' column densities.  
    \item Single-phase photoionized cloud (``Single-phase''): This model assumes a single-phase cloud with a uniform density and that the cloud is predominantly photoionized. Since all the ions are in the same photoionized phase in this model, all the observed ion species are used as the constraints on the model. 
    \item Two-phase photoionized cloud (``Two-phase''): We can think of a photoionized cloud in the line of sight with a multi-density structure \citep{stern16}. In this case, intermediate and high ions could arise from a low-density part of the gas cloud while low ions would mostly reside in the higher-density (inner) region of the cloud. All of the ion species are employed in the HMC fitting the same as the single-phase model, but there are twice as many free parameters, $\Sigma_{\rho}(N_{\rm HI, \rho}, Z_{\rho}, U_{\rho})$, where $\rho =$ high, low. 
    We note that we treat the metallicity as a free parameter in both gas phases, motivated by the possibility that they could have different metallicities due to distinct origins and mixing processes.
\end{itemize}

We present an HMC corner plot for M3 with the photoionized low-ion cloud model as an example in Fig.~\ref{fig:corner}. 
The first row of each column shows the posterior probability distribution function (PDF) of the corresponding parameter with the median and $\pm 1 \sigma$ value of the PDF. 
Fig.~\ref{fig:spaghetti} presents the comparison between this low-ion CLOUDY photoionization model and the observational constraints for both IVa (upper panel) and IVb (lower panel) detected along the M3 sight line. 
We vary each parameter over the full grid range of values in each column, keeping the other two model parameters as the best fit values (solid, colored lines). 
The vertical gray-dotted line marks the best-fit value for the given parameter on the x-axis, while the circular points with errors given by the shaded regions mark the column density measurements from the absorption-line data.

%----------------------------------------------
\section{Results} \label{sec:results}
\subsection{Photoionization Model Comparison}

%%-------------------
\begin{figure*}
    \centering
    \includegraphics[width=0.75\textwidth]{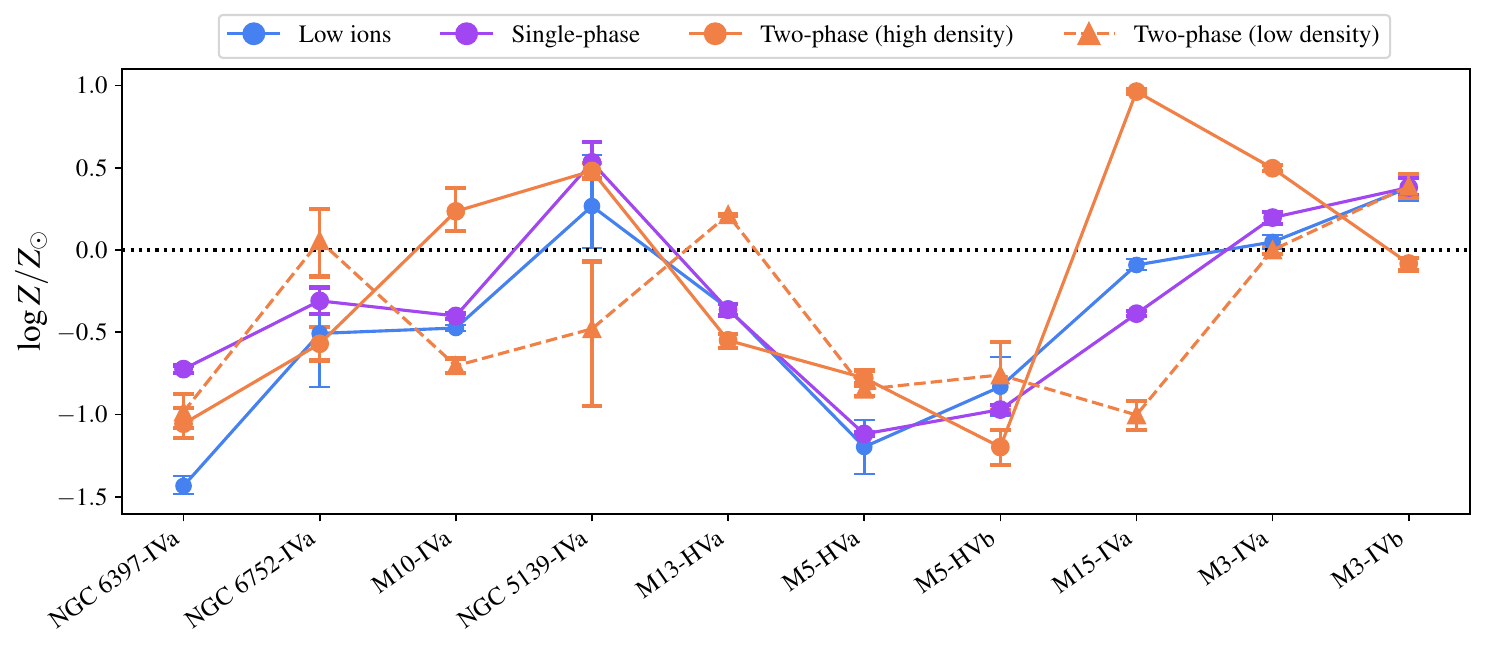}
    \caption{The metallicity comparison between photoionized cloud models. The errors are provided by the HMC PDF with $2\sigma$ confidence. The blue circle points are for ``photoionized low-ion cloud'', purple points represent ``single-phase photoionized cloud'', and orange points show ``two-phase photoionized cloud'' of which each density phase is noted with point shape and line style, circle points with solid line (high density) and triangle points with dashed line (low density). Solar metallicity $Z_\odot$ is marked with black dotted line.}
   \label{fig: Z_comparison}
\end{figure*}
%%-----------------------

 We display the metallicity of the detected DHI clouds in Fig.~\ref{fig: Z_comparison} for each of the photoionized cloud models. 
 The metallicities estimated by the ``low ion'' model are represented by blue circle points connected by the blue solid line. Similarly, the ''single-phase'' model is illustrated by purple circle points connected by the purple solid line. 
 The metallicity estimations with the ''two-phase'' model are represented as triangle points, with the high-density phase depicted in blue with a dotted line, and the low-density phase shown in orange with a dashed line. 
The metallicity estimations derived from the low ion and single-phase models generally exhibit good agreement within a few tenths of a dex. 
This result is likely attributed to the substantial contribution of low ions in the HMC fitting procedure, as they account for a significant fraction of the observed constraints in the majority of cases. 
To compare the two-phase model with the other models, the metallicity provided by the low ion model or single-phase model is compatible with that of the dominant gas phase in the two-phase model except for M15-IVa in which they converge to the middle value between the two phases. It is notable (and challenging to explain physically) that two-phase model for the  M15-IVa cloud suggests a low density gas phase with low metallicity, and a high density gas phase with solar metallicity.

%%-------------------
\begin{figure}
    \centering
    \includegraphics[width=0.48\textwidth]{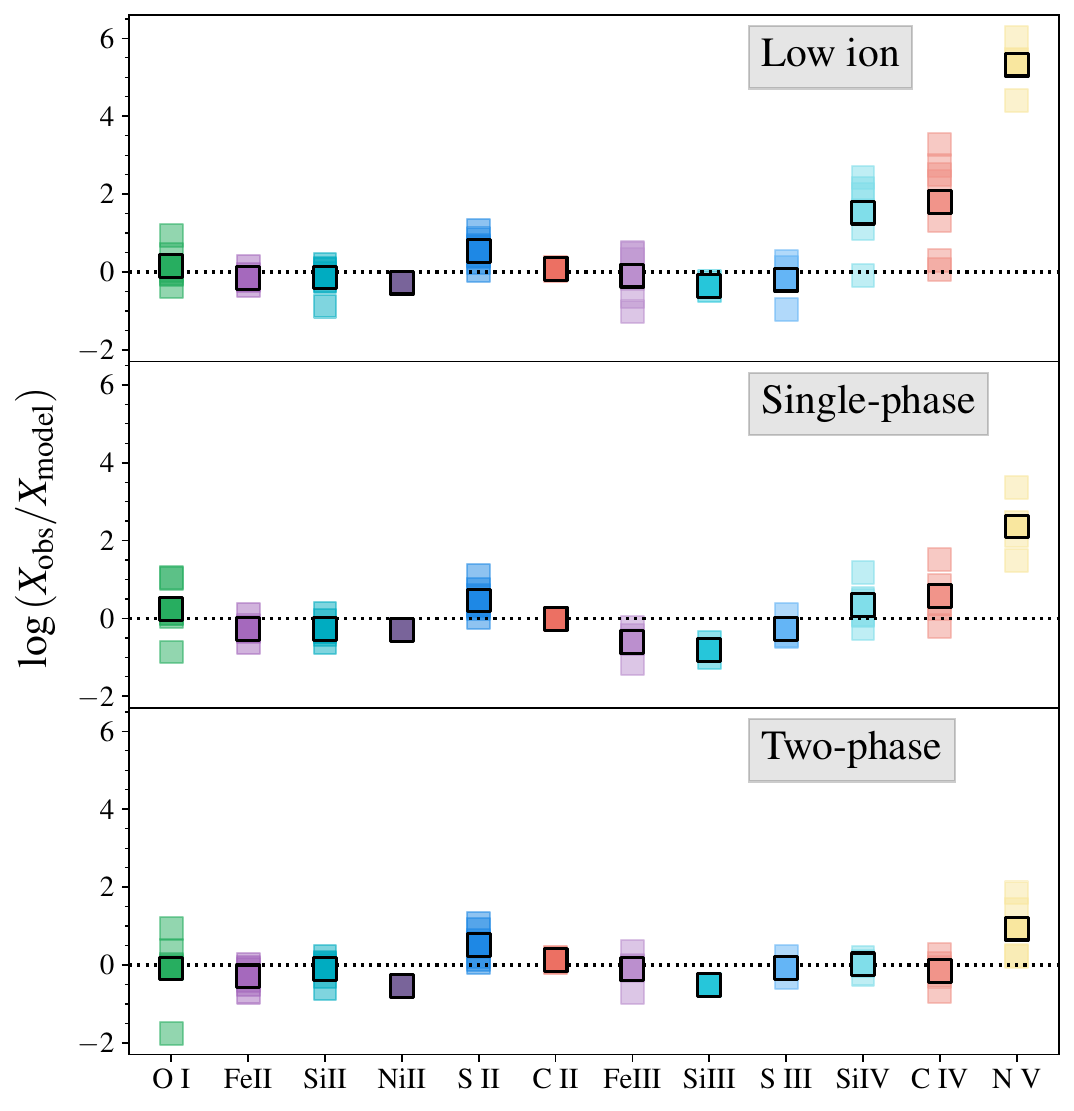}
    \caption{The deviation of the ionic column densities derived from the UV absorption observations from the values derived using best-fit Cloudy model. Ions are ordered on the x-axis in order of increasing ionization potential energy. Median deviations for every cloud are shown with a black box outline. While the low-ion model does not use the observational constraints on N$_{\rm SiIV}$, N$_{\rm CIV}$, or N$_{\rm NV}$, both the single-phase and two-phase models do.}
    \label{fig:abundance_deviation}
\end{figure}
%%-----------------------

In order to examine the goodness of fit for each model, we calculate the deviation of the observed ion column densities, $X_{\rm obs}$, from the predicted values by the best-fit model, $X_{\rm model}$. 
Fig.~\ref{fig:abundance_deviation} presents the deviation for each ion species in order of ionization potential energy. 
The scattered points along vertical lines are the deviations for each cloud. We mark the mean deviation for each ion species with the point with a black edge. 
The low ion model (top) successfully reproduces the observed low ions showing the lowest-level deviations, while the single-phase model (middle) shows larger deviations. Neither satisfactorily matches with the column density constraints from  observed intermediate ions (\ion{Si}{4}, \ion{C}{4}, and \ion{N}{5}), even these ions are considered in the HMC fitting for the single-phase model. 
This result implies that low ions and intermediate ions in the DHI gas cannot be explained simultaneously with a single-phase photoionized cloud model. Therefore, it is required to consider a multi-phase gas structure or/and other ionization mechanisms \citep{fox05, stern16, werk19}. 

The two-phase model (bottom) shows good agreement with observations, notably in successfully reproducing both low and intermediate ions.
However, we cannot conclude the two-phase model is the most suitable for DHI clouds since it requires a high-dimensional parameter space compared to the limited number of observational constraints. 
Furthermore, a thorough exploration of other potential ionization processes is essential, considering that the model still shows offsets from the data in \ion{N}{5} with about 0.8 dex. 
Thus, we discuss in detail the potential ionization mechanisms for intermediate ions in Section~\ref{subsec:intermediate}. 
In the subsequent section, our focus shifts to presenting results based on the low ion model, as it is built on the most physically robust assumptions regarding the ionization processes. Moreover, this model effectively accounts for the observed low ions and in most cases (6/8) reasonably represents the high-density gas phase metallicity of the two-phase model.

%--------------------------------
\subsection{Physical properties of the photoionized low-ion clouds} \label{subsec:low ion result}

\begin{figure}
    \centering
    \includegraphics[width=0.48\textwidth]{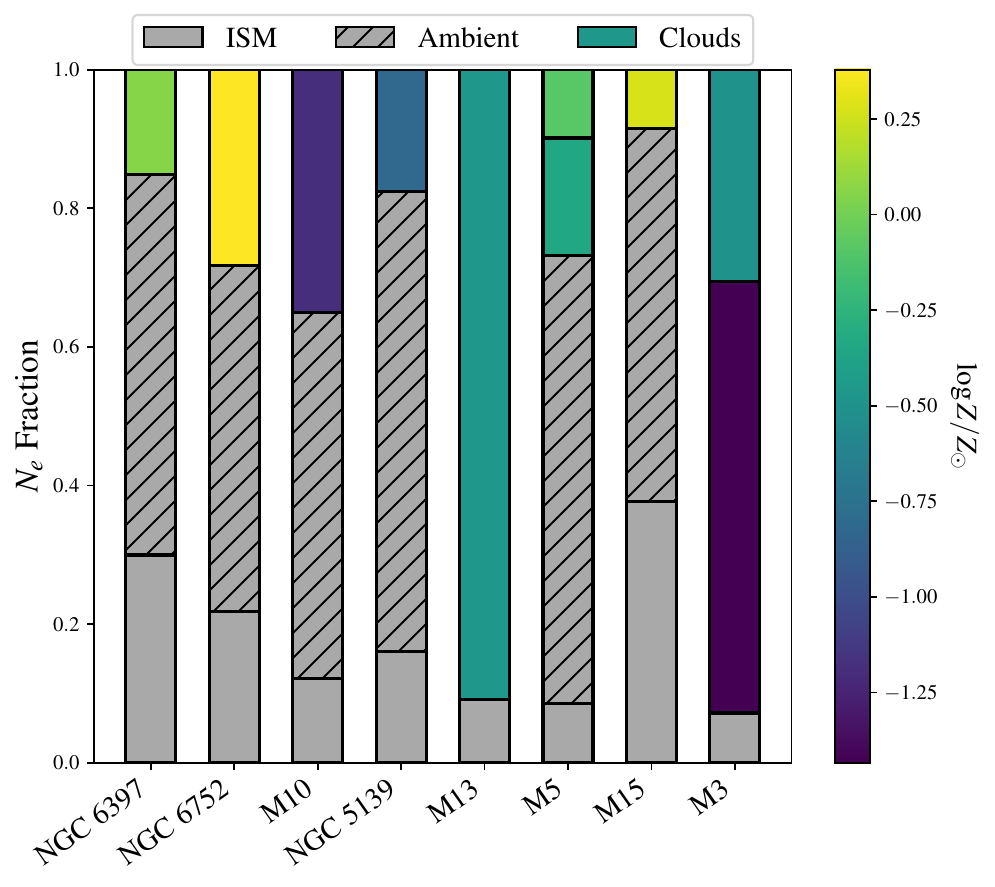}
    \caption{Bar chart showing the the electron column density, $N_e$, fraction for each gas component along the GC sight lines, ordered by increasing distance to the GCs. Each bar is divided into three components: Milky Way ISM (gray), clouds (color), and ambient (gray-hatched). The fraction of the bar representing the IVCs and HVCs is color-coded according to metallicity. The ISM fraction is calculated by the NE2001 model, as discussed in Section~\ref{sec:photoion}. The electron column density of each cloud is estimated from the HMC fitting with photoionization modeling. Since we set $N_e$ given by the pulsar DMs as an upper limit, this allows for an additional component of electron column density besides that of the clouds and ISM, possibly ambient hot halo gas. The ambient gas is the dominant source of free electrons along most lines of sight.}
    \label{fig:Ne_fraction}
\end{figure}

%------------------------
\begin{figure*}
    \centering
    \includegraphics[width=0.95\textwidth]{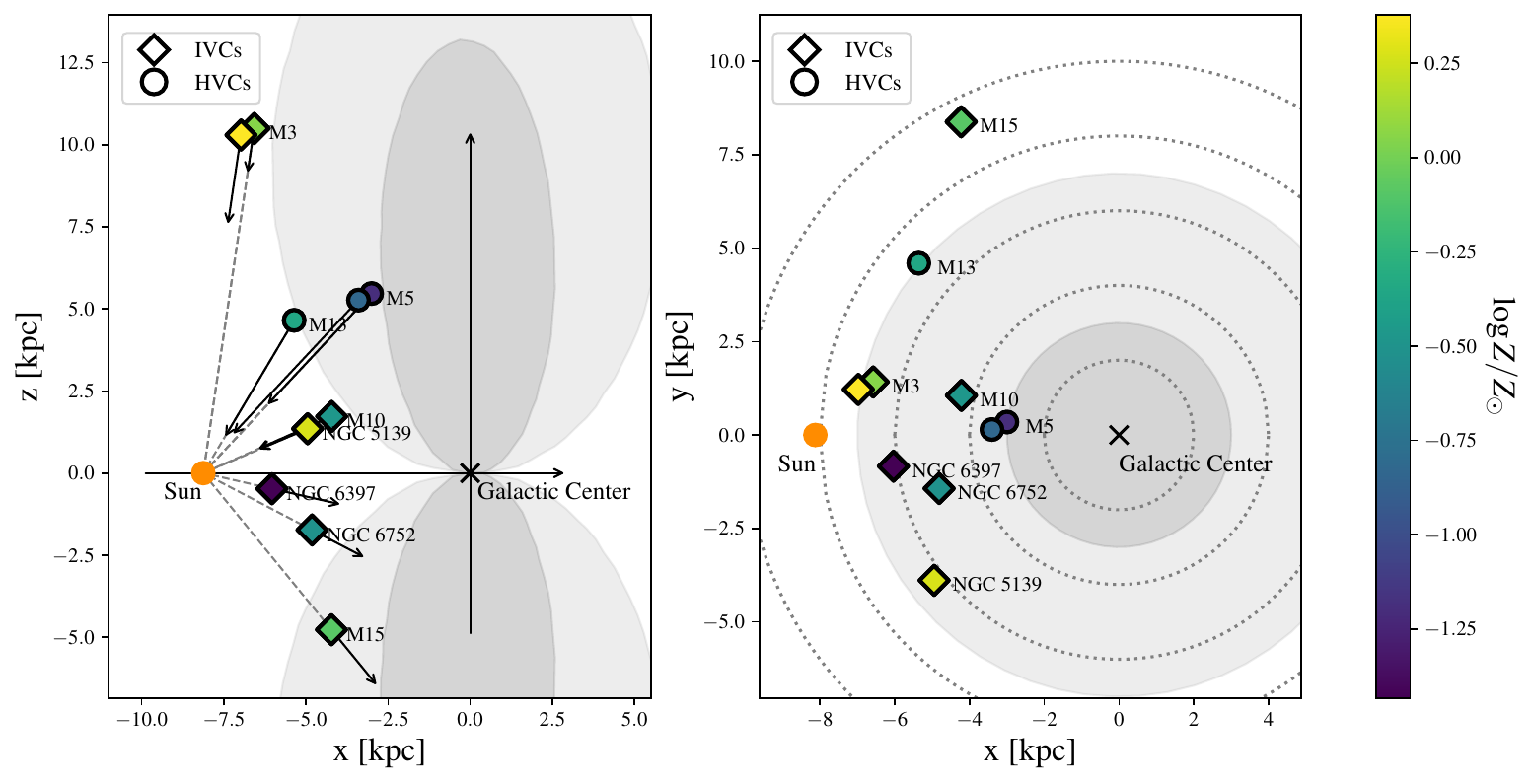}
    \caption{The detected IVCs (diamond) and HVCs (circle) in the Galactocentric system where the Sun is at (-8.122, 0, 0) kpc. The left panel shows the distribution in the XZ plane and the right panel is for the XY plane. We show our metallicity estimation in color. The vector sizes are proportional to the line of sight velocity centroids in the LSR. We also present the Fermi (dark gray-shaded) and eROSITA bubbles (gray-shaded) with their approximate size \citep{predehl20}.}
    \label{fig:gc_dist_xyz}
\end{figure*}

%------------------------------------

We present the physical properties of the detected clouds derived from HMC fitting to the photoionized low-ion cloud model in Table~\ref{tab:best-fit}. 
The \ion{H}{1} column density $N_{\rm HI}$, metallicity $Z$, and ionization parameter $U$ are the free parameters of the CLOUDY model, and the best-fit values are given by the median value of the HMC PDFs. The errors are given with $2\sigma$ confidence (95\%). 
The \ion{H}{2} column density $N_{\rm HII}$ is calculated in the model with the given best-fit parameters. 
The hydrogen number density $n_{\rm H}$ can be directly converted from the ionization parameter by $n_{\rm H} = n_{\gamma} / U = \Phi / (c U)$, where $\Phi$ is ionizing photon flux. 

%-------------------------
\begin{deluxetable*}{lccccccccc}
    \tablecaption{The cloud property estimations based on the photoionized low-ion cloud model using the HMC method. 
    The errors are given with $2 \sigma$ confidence in the PDF. 
    \label{tab:best-fit}}
    \tablehead{\colhead{GC} & \colhead{$z_{\rm max}$} & \colhead{$b$} & \colhead{Component} & \colhead{$\overline{v}_{\rm LSR}$} & \colhead{$\log{N_{\rm HI}}$} & \colhead{$\log{N_{\rm HII}}$} & \colhead{$\log{Z/Z_{\odot}}$} & \colhead{$\log{U}$} & \colhead{$\log{n_{\rm H}}$} \\
    \colhead{} & \colhead{(kpc)} & \colhead{($\degr$)} & \colhead{} & \colhead{($\rm km\ s^{-1}$)} & \colhead{($\rm cm^{-2}$)} & \colhead{($\rm cm^{-2}$)} & \colhead{} & \colhead{} & \colhead{($\rm cm^{-3}$)}
    }
    \startdata
    M3 (NGC 5272)     &  10.02 & 78.68 & IVa &  -37.3  & $19.51^{+0.05}_{-0.05}$ & 19.69 & $0.05^{+0.05}_{-0.05}$ & $-3.55^{+0.06}_{-0.06}$ & $-1.08^{+0.80}_{-0.13}$ \\
            &     &    & IVb &  -71.6  & $17.39^{+0.06}_{-0.06}$ & 19.40 & $0.38^{+0.07}_{-0.08}$ & $-2.76^{+0.10}_{-0.11}$ & $-1.87^{+0.80}_{-0.13}$ \\
    M5 (NGC 5904)     &  5.47 & 46.79 & HVa & -119.2  & $16.99^{+0.17}_{-0.17}$ & 19.18 & $-1.20^{+0.16}_{-0.16}$ & $-3.24^{+0.11}_{-0.11}$ & $-1.09^{+0.50}_{-0.21}$ \\
            &     &    & HVb & -142.9  & $17.00^{+0.20}_{-0.20}$ & 18.95 & $-0.83^{+0.18}_{-0.18}$ & $-3.45^{+0.16}_{-0.15}$ & $-0.89^{+0.50}_{-0.21}$ \\
    M10 (NGC 6254)    &  1.74 & 23.09 & IVa & -64.4   & $18.47^{+0.04}_{-0.04}$  & 19.66 & $-0.47^{+0.02}_{-0.02}$ & $-3.33^{+0.01}_{-0.01}$ & $-0.81^{+0.31}_{-0.21}$ \\
    M13 (NGC 6205)    &  4.67 & 40.94 & HVa & -106.2& $18.55^{+0.09}_{-0.07}$ & 19.92 & $-0.36^{+0.04}_{-0.04}$ & $-3.03^{+0.02}_{-0.03}$ & $-1.58^{+0.78}_{-0.07}$ \\
    M15 (NGC 7078)    &  -4.76  & -27.31 & IVa &  60.1   & $18.74^{+0.06}_{-0.07}$ & 19.22 & $-0.09^{+0.04}_{-0.03}$ & $-3.80^{+0.04}_{-0.04}$ & $-0.90^{+0.87}_{-0.17}$ \\
    NGC~5139&  1.36  & 15.05  & IVa &  -43.3  & $19.66^{+0.22}_{-0.29}$ & 19.72 & $0.27^{+0.31}_{-0.25}$ & $-3.87^{+0.22}_{-0.13}$ & $-0.30^{+0.34}_{-0.11}$ \\
            % &         & IVb &  -59.88  & $19.60^{+0.18}_{-0.20}$ & 19.81 & $0.89^{+0.10}_{-0.16}$ & $-2.67^{+0.26}_{-0.26}$ & $-1.50^{+0.34}_{-0.11}$ \\
    NGC~6397&  -0.46 & -11.94  & IVa &  51.9   & $18.46^{+0.09}_{-0.09}$  & 19.66 & $-1.43^{+0.06}_{-0.05}$ & $-3.58^{+0.04}_{-0.04}$ & $-0.24^{+0.02}_{-0.19}$ \\
    NGC~6752&  -1.72 & -25.64  & IVa &  48.0   & $18.99^{+0.41}_{-0.40}$ & 19.46 & $-0.51^{+0.29}_{-0.32}$ & $-3.72^{+0.24}_{-0.24}$ & $-0.44^{+0.33}_{-0.23}$ \\
    \enddata
    \tablecomments{The model parameters are neutral hydrogen column density \nhi, metallicity $Z/Z_{\odot}$, and ionization parameter $U$. We present ionized hydrogen column density \nhii~given by the best-fit model. }
\end{deluxetable*}
%-------------------------------------------------------------
First, we point out some sources of systematic error inherent to photoionization modeling. The primary uncertainty in our method arises from the modeled fraction of the electron column density contributed by the detected ionized clouds. Despite this uncertainty, when combined with the constraints from ion species with various ionization potentials, our DM method still yields more reliable constraints on the gas ionization parameter than photoionization modeling without the electron column constraints from the pulsar DM. When we perform the modeling without any DM constraints, there are several sight lines that are driven to much higher values of N$_{H}$ (and correspondingly lower metallicity) than allowed by the DM constraints (M3, M5, M10, and NGC~6397). For example, in the M3 sight line, the pulsar DM provides a maximum electron column of $\log{N_e} = 19.91$ (Table~\ref{tab:measurements}). Without this constraint, the best-fit model yields the total electron column density of $\log{N_e} = 21.66$, and the estimated metallicity for M3-IVb decrease to $\log{Z} = -1$, and ionization parameter increase to $\log{U} = -1.5$. The importance of the $N_e$ constraint given by pulsar DMs becomes evident considering that the metallicity estimates of ionized DHI has has been challenging since strong constraints on neutral hydrogen column density are unavailable for most sight lines. Another source of systematic error is the uncertainty in $\Phi$, which is directly proportional to the hydrogen number density. This error can be considerable because we cannot pinpoint the precise 3D location of the clouds, nor do we have precise constraints on the escape of ionizing photons from the  Milky Way (Fig.~\ref{fig:mw_radiation}). Both of these sources of error are always present for any metallicity measurements of ionized gas detected in absorption, but we cannot eliminate them even with the additional DM constraint. 

In Figure~\ref{fig:Ne_fraction}, each GC sight line is represented by a bar chart divided into three components: the Milky Way ISM (gray), clouds (colored) that correspond to the detected IVC or HVC (``Component'' in Table~\ref{tab:best-fit}), and an additional ambient medium (gray-hatched). Thus, the total electron column density given by pulsar DM, is composed of contributions from the ISM, the clouds, and the ambient medium, expressed as $N_{e,\rm DM} = N_{e, \rm ISM} + N_{e, \rm cloud} + N_{e, \rm ambient}$. The ISM fraction is calculated using the NE2001 model, while the electron column density of the clouds is derived from our HMC fitting with photoionization modeling. Given that we use $N_e$ values from pulsar DMs as an upper limit of the total electron column density, there is an uncertainty regarding the precise allocation of $N_e$ between the clouds and any remaining sources of free electrons not probed by the UV data (e.g. hot gas at T $\approx$ 10$^{6}$K). We can see that for six of our eight sight lines, the ``ambient" component accounts for approximately 50\% of the total electron column. We can additionally use the minimum and maximum $\Phi$ values  reported by \citep{fox05} to estimate a higher and lower limit of $n_{\rm H}$, respectively. Combining these two sources of systematic uncertainty, both of which affect the best-fit ionization parameters and metallicities, we conservatively estimate that they can contribute at most 0.5 dex of error on our reported values.  We conclude that our metallicities are likely generally reliable to within a few tenths of a dex. As a result, moving forward, we refer to clouds as sub-solar ($< 0.3~Z_{\odot}$), near-solar (0.3 - 1 $Z_{\odot}$), and super-solar ($> Z_{\odot}$).

The spatial distribution of the detected IVCs (diamond) and HVCs (circle) in the Milky Way is illustrated in Fig.~\ref{fig:gc_dist_xyz}. 
We mark the metallicity with color: sub-solar ($< 0.3~Z_{\odot}$; blue), near-solar (0.3 - 1 $Z_{\odot}$; purple), and super-solar ($> Z_{\odot}$; pink). 
Each range of metallicity could reflect the origin of gas clouds. Super-solar metallicity implies the Galactic origin, whereas sub-solar gas might have originated from the intergalactic medium or metal-poor satellite given that the metallicity of hot halo gas is thought to be $\sim 0.3~Z_{\odot}$ \citep{miller15}. 
The intermediate metallicity range may indicate either the mixing between metal-poor infalling gas and metal-rich Galactic medium \citep{heitsch09} or the Galactic fountain process that metal-rich gas ejected from the disk is mixed with metal-poor halo gas through condensation and falls back to the disk \citep{fraternali15, richter17}. 
Our targeted sight lines cover a wide region between the Sun and the Galactic center, and the metallicity of the DHI is largely scattered from sub-solar to super-solar. 
This wide range of metallicity may reflect the complex origin of the gas clouds and/or incomplete mixing processes in this DHI region \citep{howk18}. 
We explore the origin of gas clouds and examine possible scenarios to explain the observed metallicity distribution in Section~\ref{subsec:origin}.

The hydrogen number density spans from $\log{n_{\rm H}} = -1.4$ to $-0.2$ ($\rm cm^{-3}$), having the mean value of -0.82, and we note that this value is comparable with previous observational and simulational studies of warm ionized gas at the DHI \citep{gaensler08, melso19}. 
%corresponding to $n_{\rm H} = 0.15\ \rm cm^{-3}$. 

\begin{figure}
    \centering
    \includegraphics[width=0.4\textwidth]{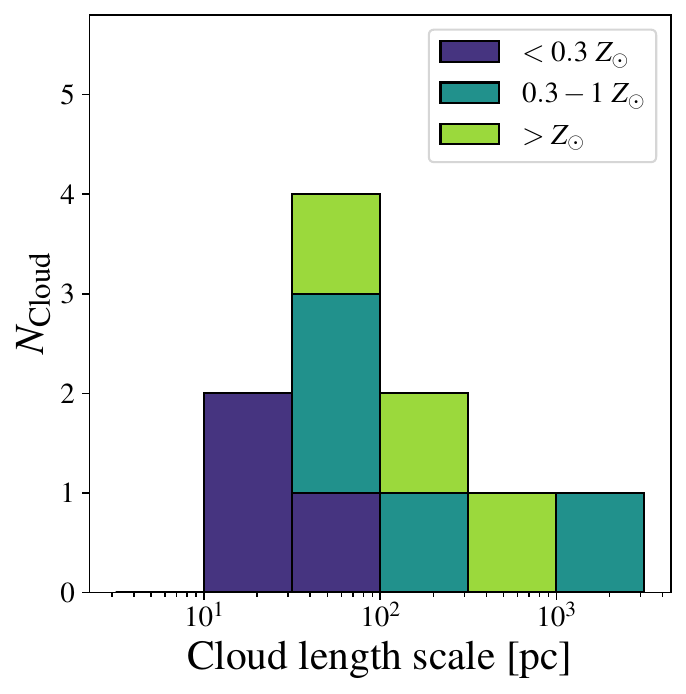}
    \caption{A histogram of cloud size. Most of the clouds have a size on the order of $\sim 100$ pc. The color shows the three metallicity ranges, less than $0.3~Z_{\odot}$ (blue), 0.3 - 1 $Z_{\odot}$ (purple), and super-solar (pink).}
    \label{fig:cloud_size}
\end{figure}

Using the hydrogen number density and the total hydrogen column density, we estimate the length scale of the gas clouds, $D = N_{\rm H} / n_{\rm H}$. 
The length scales range from 20 - 1000 pc, with a median value on the order of $\sim$ 100~pc (Fig.~\ref{fig:cloud_size}). We further note that this size scale is self-consistent with the best photoionization model in that all of the cloud sizes lie below the limit determined by the distance to the targets. 
The primary uncertainty in cloud size can be largely attributed to their undetermined locations, which dictates the flux of photoionizing radiation to which they are subjected.  Nonetheless, the cloud sizes we derive are consistent with the findings of previous observational constraints on DHI gas based on the spatial variation of \ion{Ca}{2} column density \citep{bish19}. 
In this study, the cloud sizes are determined exclusively through photoionization modeling, which provides estimates based on ion column density, gas density, and ionization states. On the other hand, in \cite{bish19}, cloud size estimation was derived from spatial investigations, assessing the extent of DHI gas clouds through observations of low-ion cloud detection and non-detection across the sky, supported by well-constrained distances. 
They assume from their study that they are tracing clouds with temperatures $T \lesssim 10^4$~K with cloud sizes $< 0.5$ kpc, similar to our constraints in this independent study using a very different cloud size estimation method.

%--------------------------
%-----------------------------------------------------

%------------------------------------

\begin{figure*}
  \centering
  \subfloat[The Galactic distribution of detected IVCs (diamond) and HVCs (circle) with an 
    all-sky \ion{H}{1} velocity map from the 21-cm HI4PI survey \citep{westmeier18}. 
    Our clouds are potentially part of the known IVC and HVC complexes, Complex C, Complex K, and Complex gp. The three complexes are presented with boxed outlines having a color corresponding to their average LSR velocity.]{\centering
    \includegraphics[width=\linewidth]{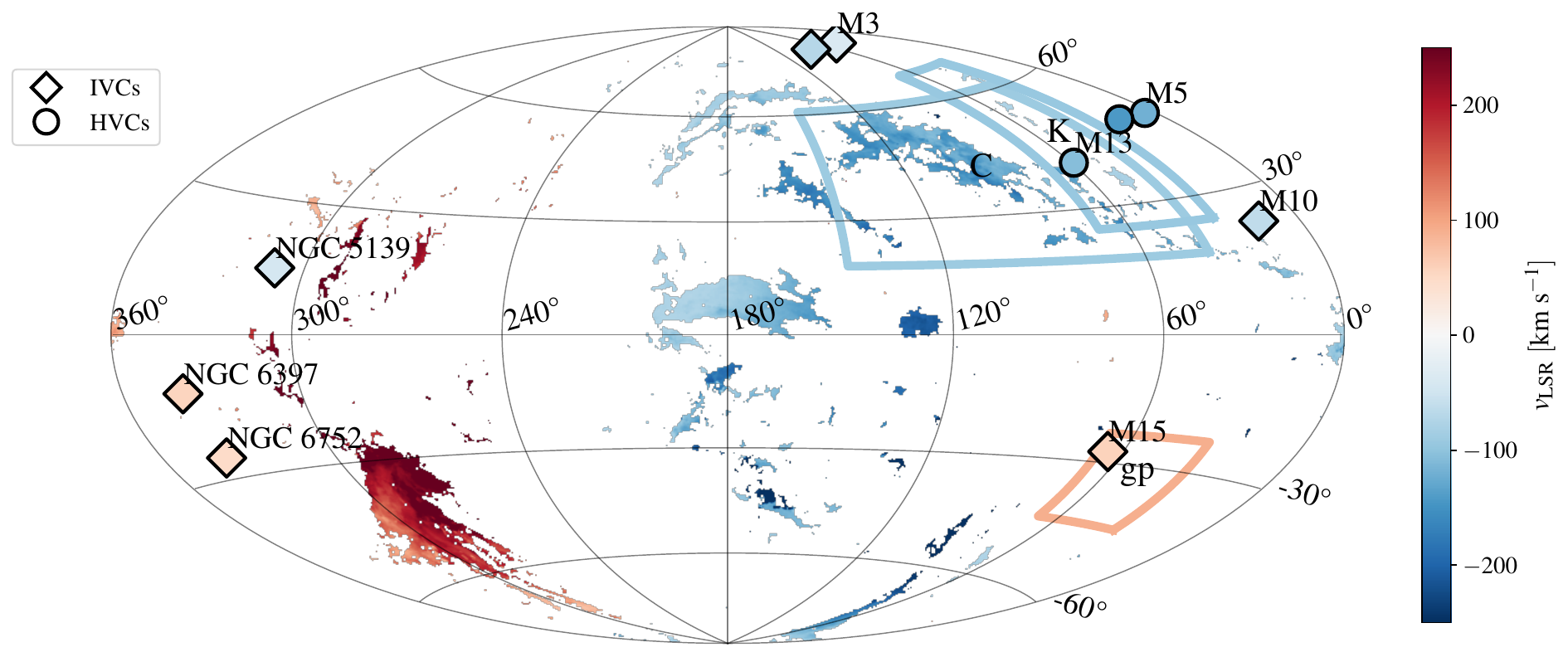}
    \label{fig:gc_dist}}
  \quad
  \subfloat[The WHAM map for Complex C (left), Complex K (center), and Complex gp (right). The black boxes correspond to the location of the known complexes. The H$\alpha$ intensity is integrated for the LSR velocity ranges of each complex, presented on top of the map.  The LSR velocity ranges are referred to \cite{vanwoerden04}. 
  We note that the WHAM survey covers the velocity range of roughly $(-100, 100)\ \rm km\ s^{-1}$, 
  thus it only covers a small fraction of the velocity range of Complex~C. 
  The range of intensities varies for each map, as indicated by the corresponding color bar. The H~I map from the HI4PI survey \citep{hi4pi16} is presented with blue contours with the column density of $N_{\rm HI} = 10^{19.5}, 10^{19}$, and $10^{18}\ \rm cm^{-2}$ for Complex C, Complex K, and Complex gp, respectively. 
  We show the location of our three sight lines that are potentially associated with the complexes (magenta stars).]{\centering
    \includegraphics[width=\linewidth]{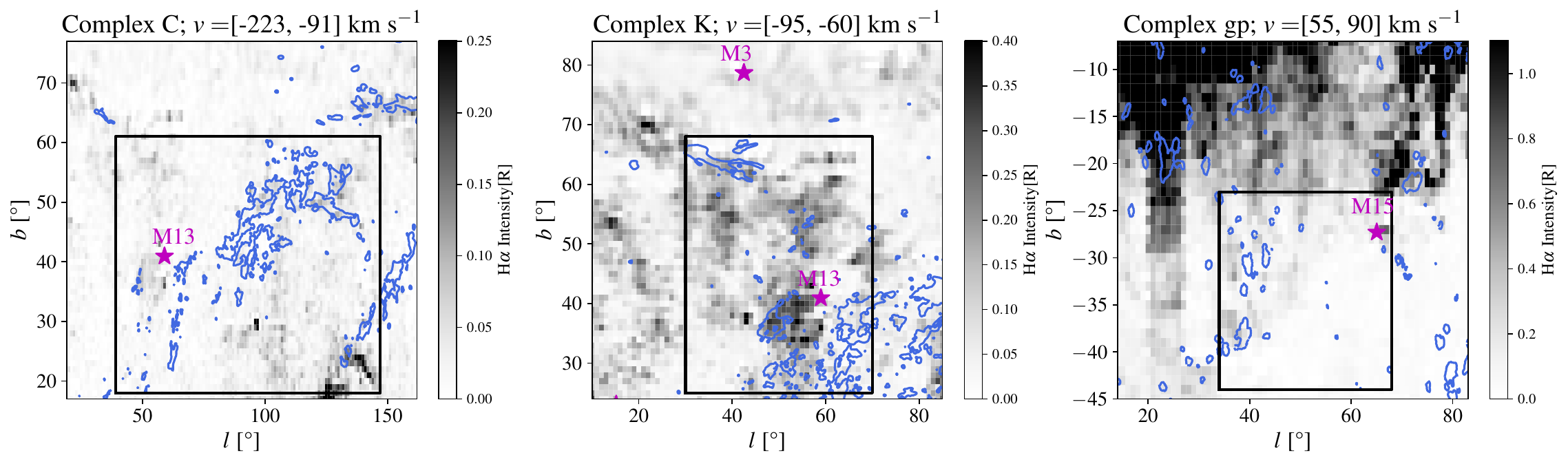}
    \label{fig:wham_map}}
  \caption{(a) The Galactic distribution of detected IVCs and HVCs with all-sky H~I map. (b) The WHAM map of the known HVC and IVC complexes that are potentially associated with the detected ionized clouds.}
  \label{fig:complexes_map}
\end{figure*}

%------------------------------------------------------------------------
\section{Discussion} \label{sec:discussion}

\subsection{Photoionization of Intermediate Ions} \label{subsec:intermediate}
We present the physical properties of the identified ionized DHI clouds in Section~\ref{subsec:low ion result}, with a specific focus on metallicity based on detailed photoionization modeling of low ions. 
The results exhibit good agreement with observed low ions and maintain consistency with pulsar DMs. 
The disparity between measured column densities of intermediate ions, \ion{Si}{4}, \ion{C}{4}, and \ion{N}{5}, and their predicted values from the low ion and single-phase model indicates that these ions likely reside in distinct gas phases (top and middle of Fig.~\ref{fig:abundance_deviation}). 
We here explore the potential photoionization of intermediate ions and discuss its validity. 
We discuss other potential ionization mechanisms for the intermediate ions in Appendix~\ref{subsubsec:collisional}.

We conduct the ``two-phase" photoionized cloud model in Section~\ref{sec:photoion} with the motivation of the hierarchical structure of photoionized clouds along the line of sight, where small dense clouds are embedded in larger low-density clouds \citep{stern16}. 
The results presented in Fig.~\ref{fig:abundance_deviation} (bottom) demonstrate that the observed column densities of \ion{Si}{4} and \ion{C}{4} can be reproduced by the photoionization model, in particular, a significant fraction of \ion{Si}{4} and \ion{C}{4} resides in the low-density cloud at a temperature of $T \sim 10^4$~K.

The \nhii\ of low-density clouds is approximately 70~\% on average compared to that of high-density clouds, indicating that low-density clouds contribute significantly to the total \nhii\ in the two-phase photoionization model. 
We also find a discrepancy in the metallicity estimation for each density phase (Fig.~\ref{fig: Z_comparison}). 
This discrepancy may reflect the nature of the DHI where multi-phase gas flows with various origins are incorporated and affected by feedback, possibly being accreted onto the disk and fueling star formation. 
The distance to the background halo star serves as a strict upper limit for the cloud length scale. 
By using this upper limit, we can validate the photoionization model and rule it out if it predicts a larger size for the clouds than the distance to the star. 
The length scales of the low-density cloud fall within the distance to the star, with a mean value of $\simeq 1$~kpc. 
Therefore, from the cloud size analysis alone, there remains a possibility that \ion{Si}{4} and \ion{C}{4} originate from photoionized low-density gas clouds. 
The low-density clouds in the best-fit two-phase model have a median hydrogen number density of $n_{\rm H} = 0.01~\rm cm^{-3}$, while that of the high-density phase gas is $n_{\rm H} = 0.2~\rm cm^{-3}$. 
Considering a similar gas temperature of the two phases, $T \sim 10^4$~K, there is more than an order of pressure imbalance between the two phases under the hydrostatic equilibrium assumption. Consistent with \cite{stern16}, we conclude that either there must be a hydrodynamic solution to explain this density structure or the highly ionized species are collisionally ionized as discussed in Appendix~\ref{subsubsec:collisional}.

\subsection{Association with Known \ion{H}{1} IVCs and HVCs} \label{subsec:known}
In Fig.~\ref{fig:gc_dist}, we present the Galactic distribution of the identified IVCs (diamonds) and HVCs (circles) in this work with their LSR velocities, along with an all-sky \ion{H}{1} velocity map sourced from the 21~cm HI4PI survey \citep{westmeier18}. 
The ionized clouds identified in this work could potentially be part of or associated with known \ion{H}{1} IVCs and HVCs \citep{wakker01}. 
This potential association is based on their spatial alignment in Galactic coordinates and comparable LSR velocities. 
The potentially associated IVC and HVC complexes are marked with boxed contours colored by their LSR velocities. Their locations in Galactic coordinates are referenced from \cite{vanwoerden04}. 
Fig.~\ref{fig:wham_map} shows a zoom-in H$\alpha$ map of the complexes, and the black boxes correspond to the colored boxes in Fig.~\ref{fig:gc_dist}. 
The H$\alpha$ map is sourced from the Wisconsin H-Alpha Mapper (WHAM) survey \citep{haffner03}, with the H$\alpha$ intensity integrated over the LSR velocity range of the known complexes \citep{wakker01}. 
We overplot an H~I contour (blue) as well with the column density of $N_{\rm HI} = 10^{19.5}$, $10^{19}$, and $10^{18}\ \rm cm^{-2}$ for Complex C, Complex K, and Complex gp, respectively.

\subsubsection{M3 - Complex K} \label{subsec:M3}
Two IVCs, labeled IVa and IVb, are identified along the line of sight toward M3, exhibiting $v_{\rm LSR}$ of $-37\ \rm km\ s^{-1}$ and $-72\ \rm km\ s^{-1}$, respectively (see Fig.~\ref{fig:vpf} and Table~\ref{tab:best-fit}). 
Their negative LSR velocities suggest motion indicative of an infalling gas component, particularly notable since M3 is situated near the galactic North Pole ($b = 79 \degr$), where the rotational component of the motion of the Milky Way disk is negligible. 
It is noteworthy that IVa probably traces IV Arch, a huge \ion{H}{1} complex. It is evident from its high \ion{H}{1} column density of $N_{\rm HI} \sim 10^{19.5}\ \rm cm^{-2}$ and its solar metallicity (Table~\ref{tab:best-fit}) that is consistent with previous estimates \citep{wakker01}. 
On the other hand, for IVb, we suggest the possibility of its association with Complex K, an IVC complex characterized by LSR velocities in the range of $-90 < v_{\rm LSR} < -60\ \rm km\ s^{-1}$.

The presence of the IVb cloud was initially reported by \cite{deboer84} based on \ion{C}{2} and \ion{C}{4} absorption features observed at $v_{\rm LSR} \sim -70\ \rm km\ s^{-1}$ in the UV spectrum of M3 vZ1128 (NGC5272-ZNG1) taken by the International Ultraviolet Explorer (IUE). 
\cite{wakker01} discussed the potential association of this cloud with Complex K; however, they contended that the detection is inconclusive due to the absence of strong absorption lines in the FUSE spectrum of the star and the lack of associated emission in the \ion{H}{1} 21~cm map surrounding it. 
It is crucial to highlight that the high S/N COS spectrum of the same star, combined with joint Voigt profile fitting employed in this study, enables us to differentiate IVb from the strong absorption features of IVa or the Milky Way ISM. 
The low \ion{H}{1} column density estimated at $N_{\rm HI} \sim 10^{17.4}\ \rm cm^{-2}$ through the HMC fitting is consistent with non-detection in \ion{H}{1} 21~cm emission. 
If IVb is indeed associated with Complex K, it indicates the presence of a more extensive ionized structure surrounding the complex.
It is supported by the WHAM map near Complex K (center; Fig.~\ref{fig:wham_map}). H$\alpha$ emission around the neutral complex extends to the M3 sight line. 
Furthermore, the super-solar metallicity of IVb, measured at $Z = 2.4\ Z_{\odot}$, implies a likely origin as a Galactic fountain-driven cloud. 
We note that this metallicity value is higher than previous measurements for neutral part of Complex~K ($Z \sim 1-2\ Z_{\odot}$; \citealt{wakker01, vanwoerden04}).

\subsubsection{M13 - Complex C}
We detect an ionized HVC at $v_{\rm LSR} = -106\ \rm km\ s^{-1}$ along the line of sight toward M13 that is passing through a lower latitude part of Complex C, a well-known massive, infalling HVC. 
Although M13 is closer than the average distance to Complex C ($d \sim 10$~kpc; \citealt{thom08}), we cannot rule out the possibility that the detected HVC is part of a more extended ionized region of the complex \citep{fox04}.

Complex C displays low metallicity across the complex in the range of $Z = 0.1-0.3\ Z_{\odot}$ \citep{tripp03, fox23} 
consistent with our metallicity estimation of $Z = 0.4\ Z_{\odot}$. 
It could be possible that the detected HVC lies within the region where metal-poor HVC interacts with a fountain outflow \citep{tripp03}. 
Another potential scenario is that Complex C is formed via the Galactic fountain  \citep{fraternali15}, and the detected ionized HVC traces a more metal-rich part of the fountain-driven cloud. 
It is noteworthy that the hydrodynamic simulation of the fountain-driven formation of Complex C by \cite{fraternali15} shows a slight metallicity gradient by \ion{H}{1} column density, having higher metallicity at lower $N_{\rm HI}$. 
Our metallicity estimation is in agreement with the predicted value at $N_{\rm HI} \sim 10^{18.5}\ \rm cm^{-2}$.

\subsubsection{M15 - Complex gp}
The ionized IVC in the sight line toward M15 is part of a known IVC, Complex gp, with LSR velocity of $v_{\rm LSR} \sim 60\ \rm km\ s^{-1}$. 
The metallicity of Complex gp has been thought to be near-solar based on equivalent widths of several UV absorption lines \citep{wakker01}, which is consistent with our measurement of $Z = 0.8\ Z_{\odot}$.

\subsubsection{M5 - Fermi Bubbles} \label{subsubsec:M5}
The HVCs toward M5 have been regarded to be embedded in the Fermi bubbles (Fig.~\ref{fig:gc_dist_xyz}). The Fermi bubbles are largely extended ($\sim -20\degr < l < 20\degr$ and $-50\degr < b < 50\degr$) plasma bubbles launched at the Galactic center \citep{su10}, and they are clear evidence of Galactic outflows. 
Several HVCs tracing these outflows were detected within the bubbles, and \cite{ashley22} showed that they 
have a wide range of metallicity from below $0.2\ Z_{\odot}$ to $3.2\ Z_{\odot}$. 
They explained this wide scatter in metallicity by two populations of clouds with different origins. 
One population of clouds has a Galactic origin, similar to the bubbles themselves, so they are likely to be metal-rich and found at low Galactic latitudes. 
The other population is pre-existing low-metallicity halo gas shocked and accelerated by the Fermi bubbles as they expand. 
This latter population can explain the Fermi bubble HVCs with very low metallicity below $\sim 0.2\ Z_{\odot}$, and is more likely to be found at high latitudes. 

The HVCs that we detect toward M5 with sub-solar metallicity of $\sim 0.1\ Z_{\odot}$ here were also studied in \cite{ashley22}; however, they regarded them as super-solar clouds with the metallicity of $3.2\ Z_{\odot}$, referring to a previous study (\citealt{zech08}, hereafter Z08). 
We describe a few possible reasons that might have given rise to such a large discrepancy. 
First, our study considers the \nhi~measured with a \textit{FUSE} spectrum as a lower limit (\nhi $> 10^{16.46}\ \rm cm^{-2}$) because of the visible saturation. 
 Z08 used $10^{16.50}\ \rm cm^{-2}$ of \nhi~as the actual measurements of \nhi, and consequently, this value is lower than the \nhi~estimated by our photoionization modeling of $N_{\rm HI} =  10^{17.30}~\rm cm^{-2}$. 
In addition, while Z08 used the Legendre polynomial to fit the continuum, we normalize the spectra with a stellar-like continuum, characterized by a blackbody and flat in a narrow wavelength range, 
to avoid an over-correction. 
Last, and most importantly, they derived a metallicity only using \ion{O}{1} and performed an ionization correction 
using silicon ions at different ionization stages (\ion{Si}{2}, \ion{Si}{3}, and \ion{Si}{4}). 
However, the \ion{Si}{3} absorption line is saturated as they note, therefore, it could affect the ionization correction and metallicity.

If we assume that the HVCs detected toward M5 have low metallicity as we derive, 
a potential explanation for their low metallicity would be that they originated from low-metallicity halo gas and were accelerated by the Fermi bubbles as discussed in \cite{ashley22}.

\subsection{The Nature of the DHI Gas Clouds} \label{subsec:origin}
%---------------------

%-----------------------------
We find that ionized DHI gas at $|z| < 10$~kpc has a wide range of metallicities, $0.04 - 3\ Z_{\odot}$. 
This spread suggests that ample inflows and outflows from multiple origins are incorporated in the DHI region.

Historically, the categorization of IVCs and HVCs, primarily based on their LSR velocities, has treated them as distinct populations due to their differing metallicities. 
IVCs, distributed within $d < 1-2$~kpc, exhibit near-solar metallicity, suggesting a Galactic fountain origin. 
On the other hand, HVCs, found at greater distances in the halo (mostly $d < 10-15$~kpc), show lower metallicites, indicating an external origin, possibly from satellites or the IGM. 
Utilizing a significant sample of IVCs and HVCs detected in UV spectra of halo stars, covering a wide fraction of the sky \citep{lehner22}, \cite{marasco22} challenged this division. 
They proposed that the kinematics of ionized IVCs and HVCs likely constitute a unified population characterized by a combination of diffuse inflows and collimated outflows, rather than separate populations, thereby supporting the Galactic fountain concept.

At the same time, the region near the Galactic center where the samples studied here mainly cover exhibits dramatic and complex outflow features. 
As introduced in Section~\ref{subsubsec:M5}, the Fermi/eROSITA bubbles are extended to several kpc, and HVCs tracing these outflows have been detected in multiphase \citep{diteodoro18, diteodoro20, ashley22}. 
Furthermore, \cite{mou23} argued that the asymmetric structures in morphology and surface brightness shown in the eROSITA bubbles can be reproduced with the CGM wind model, implying dynamic gas flows around the Milky Way. 

\begin{figure*}
    \centering
    \includegraphics[width=0.85\textwidth]{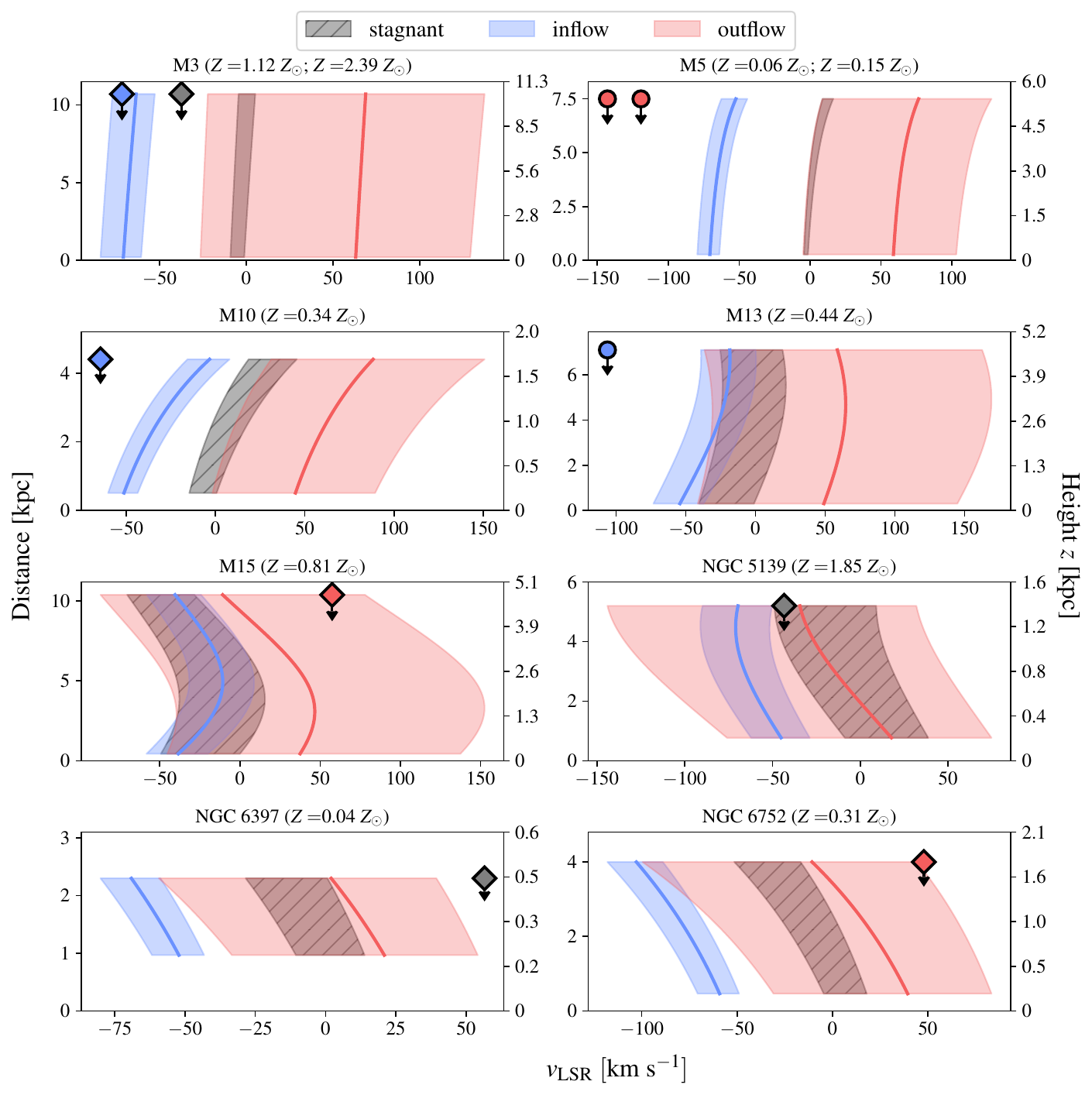}
    \caption{Kinematic modeling of LSR velocity for each sight line and observed LSR velocity of the detected IVCs (diamond) and HVCs (circle). LSR velocity is a function of distance from the Sun (left y-axis), and the maximum distance is given by the distance to the targeted halo stars. We present the predictions of inflow (blue-shaded), outflow (red-shaded), and stagnant (hatched) models with the parameters given in Table~\ref{tab:kinematic}. The color of the data points denotes the categorization of inflow (blue), outflow (red), and ambiguous (gray).}
    \label{fig:time_metal}
\end{figure*}

In this section, we compare our samples with several kinematic models and discuss the nature of DHI gas clouds. 
We apply the rotation-dominated inflow and outflow models suggested by \cite{marasco22} that consider disk-like DHI gas of which velocity vectors are in cylindrical coordinates, $(v_{\rm r}, v_{\phi}, v_{\rm z})$. 
The kinematic models provide the LSR velocity of a gas cloud at a specific distance from the Sun for any sight lines in the sky. 
Since we can only determine the upper limit of the distances to the detected DHI clouds from background halo stars, we calculate a range of acceptable LSR velocities for our sample of sight lines as a function of distance. We then compare these predictions with the observed LSR velocities (Fig.~\ref{fig:time_metal}).
For reference, we also consider stagnant models where only the rotational motion exists, thus $v_{\rm r} = v_{\rm z} = 0\ \rm km\ s^{-1}$. 
The applied model parameters are presented in Table~\ref{tab:kinematic}. The parameters for the inflow and outflow models are directly adopted from the best-fit parameters provided by \cite{marasco22}. 

%---------------------------------------
%---------------------
\begin{deluxetable}{lcccc}
    \tablecaption{Kinematic model parameters.  
    \label{tab:kinematic}}
    \tablehead{\colhead{Parameter} & \colhead{Inflow} & \colhead{Outflow} & \colhead{Stagnant} \\
    \colhead{($\rm km\ s^{-1}$)} & \colhead{} & \colhead{} & \colhead{}}
    \startdata
    $v_{\rm r}$ & $30^{+15}_{-11}$   & $-22^{+39}_{-35}$  & 0                  &  \\
    $v_{\phi}$  & $233 \pm 18$       & $232^{+91}_{-54}$  & $200 \pm 30$       &  \\
    $v_{\rm z}$ & $-67^{+8}_{-11}$   & $-22^{+39}_{-35}$  & 0                  & 
    \enddata
\end{deluxetable}

%-----------------------------
%---------------------
\begin{figure}
    \centering
    \includegraphics[width=0.48\textwidth]{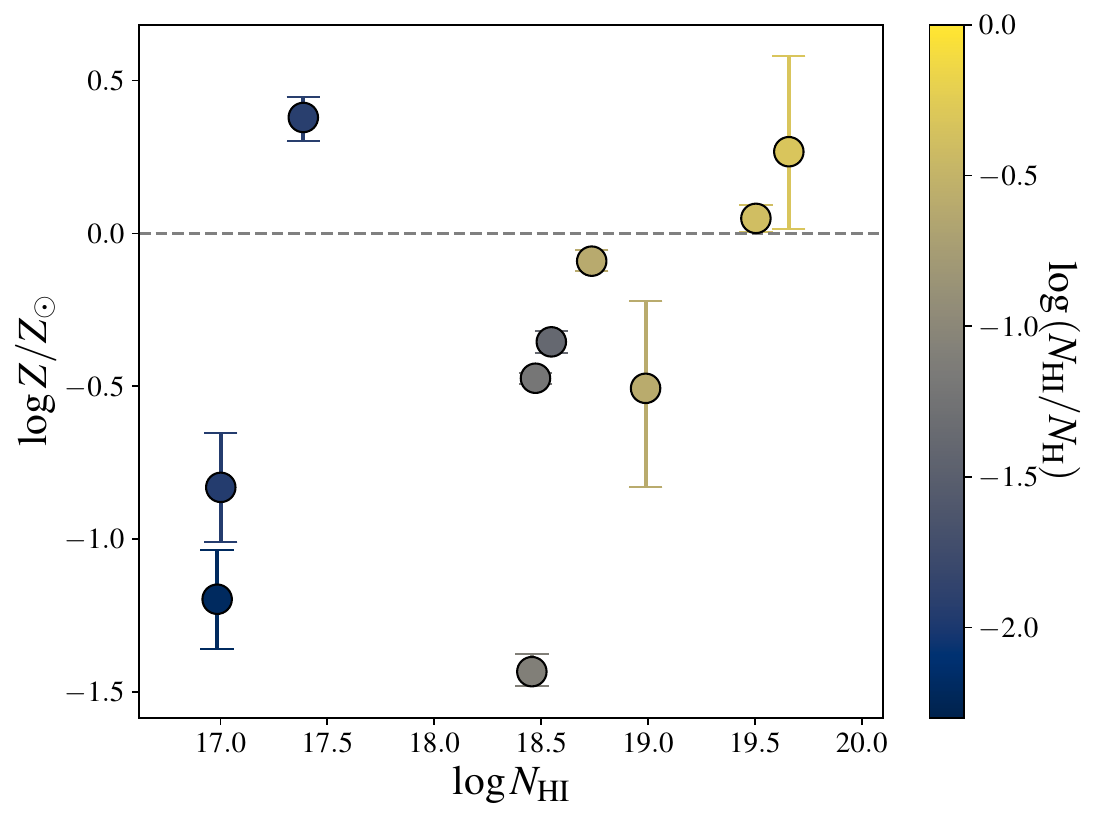}
    \caption{The estimated \nhi\ and metallicity of the ionized DHI clouds. The color shows \ion{H}{1} fraction over the total hydrogen. We note that most clouds are more than 90~\% ionized, and the existence of metal-poor gas clouds in the form of ionized clouds.}
    \label{fig:HI_to_Z}
\end{figure}
%---------------------------------------

In Fig.~\ref{fig:time_metal}, we classify the detected IVCs (diamond) and HVCs (circle) into three groups, inflow (blue), outflow (red), and ambiguous (gray), based on the comparison with the kinematic models. The exceptions are the HVCs in the sight line of M5. Despite their negative LSR velocities, they are likely to trace the outflows as discussed in Section~\ref{subsubsec:M5}. 
\cite{diteodoro18} proposed an outflow model where gas moves away from the Galactic center at a constant velocity. 
This model adequately explains the negative LSR velocities observed for the M5 HVCs by suggesting that their motion is more horizontal than vertical, implying their relative proximity, approximately 3-4~kpc away. 
In addition, we note that the HVC toward M13 could potentially be part of Complex C, a well-known infalling HVC.

In order to further examine the nature of these gas clouds, we calculate the inflow and outflow timescale $\tau$ for infalling and outflowing gas, respectively. 
We assume that the vertical velocity component, $v_z$, is solely responsible for the difference between the measured $v_{\rm LSR}$ and LSR velocities given by the stagnant model, $v_{\rm stag}$. Based on this assumption, we define $\tau$ as follows. 

% \begin{equation}
%     \tau \equiv \bigg | \frac{z_{\rm max}\sin{|b|}}{v_{\rm LSR}-v_{\rm stag}} \bigg | 
%     \approx 10 \left( \frac{z_{\rm max}\sin{|b|}}{\rm kpc}\right) \left( \frac{100\ \rm km\ s^{-1}}{v_{\rm LSR}-v_{\rm stag}}\right) \rm Myr,
% \end{equation}

\begin{align}
    v_z &= \frac{v_{\rm LSR} - v_{\rm stag}}{\sin |b|}, \label{eq:v_z}\\
    \tau &\equiv \frac{z_{\rm max}}{|v_z|} 
    = \left| \frac{z_{\rm max} \sin |b|}{v_{\rm LSR} - v_{\rm stag}} \right| \nonumber \\
         &\approx 10 \left( \frac{z_{\rm max}}{1\ \rm kpc} \right) \left( \frac{\sin |b|}{1} \right) \left( \frac{100\ \rm km\ s^{-1}}{v_{\rm LSR} - v_{\rm stag}} \right)\ \rm Myr, \label{eq:tau}
\end{align}

where $z_{\rm max}$ is the maximum Galactic height, and $b$ is Galactic latitude. 
The characteristic timescales for inflows and outflows mostly range between 6-35 Myrs, except for M3-IVb, which has a timescale of about 150~Myrs. We note that our estimates of inflow/outflow timescales approximate the duration required for gas accretion or the time elapsed since gas clouds were ejected.  

For M3-IVb, this longer timescale is attributed to its location at a higher Galactic height of 10~kpc, compared to other targeted GCs predominantly within 5~kpc.
We discuss the potential association of M3-IVb with Complex K in Section~\ref{subsec:M3}. 
If we adopt the z-height limit of Complex K from \cite{wakker01} ($z < 4.5$~kpc), the timescale falls to around 65 Myrs. 
Given its LSR velocity of $v_{\rm LSR} \simeq -70\ \rm km\ s^{-1}$ and high Galactic latitude ($b = 79\degr$), we infer that the cloud is indeed infalling into the disk, with a high likelihood of accreting onto the star-forming disk within this timescale. 
Its super-solar metallicity of $Z = 3.2\ Z_{\odot}$ suggests a Galactic origin, indicative of it being a part of a Galactic fountain. 
In addition, previous studies of gas flows near Complex K for both cool \citep{bish19} and warm phases \citep{werk19} showed that the inflows are confined within a z-height of 3.4~kpc, and their physical and kinematic properties further support a Galactic fountain model.

Unlike M3-IVb, we also observe metal-poor infalling clouds in the sight lines of M10 and M13 (Fig~\ref{fig:time_metal}). 
Given their low Galactic height of $z < 10$~kpc, these clouds are likely to survive and accrete onto the disk \citep{tan23}. 
Both metal-poor and metal-rich outflows are also observed. We propose that metal-rich clouds trace outflowing material, while metal-poor clouds trace halo gas that has been shocked and accelerated by outflows, as discussed in Section~\ref{subsubsec:M5}. 
This observation highlights the complexity of gas flows on the small scale of the DHI. 
The presence of both metal-rich and metal-poor components in inflows and outflows indicates that metallicity alone cannot reliably trace the nature or origin of gas flows.

Notably, we detect low-metallicity ionized clouds with $N_{\rm HI}< 10^{19}~\rm cm^{-2}$, which is challenging to detect and determine metallicity given the \ion{H}{1} 21~cm emission detection limit for many surveys (Fig.~\ref{fig:HI_to_Z}). 
The Galactic chemical evolution model requires the continuous accretion of low-metallicity gas to explain the observed stellar abundance and scaling laws \citep{chiappini01, sanchez14}, and lines of hydrodynamic simulation have predicted that a substantial portion of gas accretion occurring in the form of ionized matter (e.g., \citealt{heitsch09, joung12, kim18}). 
The detection of low-metallicity ionized clouds evidently shows gas accretion in the form of ionized material and implies its role in fueling Galactic star formation.

\section{Summary} \label{sec:summary}
In this work, we estimate the metallicity of the diffuse warm ionized gas at the DHI with a strong constraint on ionized hydrogen column density given by the radio pulsar DM. 
We analyze archival COS, STIS, and \textit{FUSE} spectra of 10 halo GCs and detect ionized IVCs and HVCs along the 8 sight lines within the Galactic height of $z \lesssim 10$~kpc.

All the absorption components within $|v_{\rm LSR}| < 200\ \rm km~\rm s^{-1}$ are identified, therefore, 
the ionized clouds with $|v_{\rm LSR}| > 20\ \rm km~\rm s^{-1}$ are deblended from the Milky Way ISM absorption and stellar absorption. 
We measure their ion column densities separately by Voigt profile fitting to estimate the metallicity of an individual gas cloud. 

We perform photoionization modeling using CLOUDY under three distinct scenarios: 
(1) where low ions exist in a photoionized gas phase while intermediate and high ions are collisionally ionized (referred to as the ``low ion” model), 
(2) where all ions are uniformly photoionized within a single gas phase (``single-phase”), 
and (3) where all ions are photoionized, but the clouds are structured into two distinct density phases (``two-phase”). 
We find that low ions and intermediate and high ions cannot be in a single phase of photoionized gas, clearly indicating the multi-phase structure of ionized clouds. 

We present self-consistent and statistically preferred metallicity estimates of ionized IVCs and HVCs using the HMC method based on the low ion model. The DHI gas clouds show a large scatter in metallicity, spanning the range of $0.04- 3\ Z_{\odot}$. These inhomogeneities may indicate a widespread Galactic Fountain in which outflowing and inflowing gas are neither well-mixed nor of a single origin. 

Some of our samples may trace the extended ionized envelope of known neutral HVC and IVC complexes. The IVC detected in the sight line of M3 may be part of the ionized envelope of Complex K, and its super-solar metallicity of $3.2\ Z_{\odot}$ and the inflow signature suggest a fountain-driven flow. The M13 HVC might be associated with Complex C, a well-known infalling low-metallicity HVC, and our metallicity estimate based on low ions is $Z = 0.4\ Z_{\odot}$. This value is slightly higher than previous measurements of the neutral parts, possibly indicating mixing with metal-rich Galactic material. The IVC in the sight line of M10 could be an ionized counterpart of Complex gp, and its near-solar metallicity of $Z = 0.8\ Z_{\odot}$ is comparable to the metallicity estimate of the neutral part. We also report low-metallicity HVCs toward M5 that are potentially associated with the Fermi bubbles. Interestingly, these findings diverge from previous studies that suggested a super-solar metallicity of $3.2\ Z_{\odot}$ for these HVCs \citep{zech08, ashley22}. Although they trace outflows, their low metallicity of $\sim 0.1\ Z_{\odot}$ suggests that they are likely preexisting halo gas accelerated by the expansion of the bubbles rather than directly tracing outflow material. 

In the unambiguous cases for which we are able to identify inflows and outflows, the characteristic timescales on which material cycles into and out of the disk mostly range between $6-35$ Myrs. We detect a super-solar infalling cloud, and a low-metallity outflowing cloud, which poses a challenge for Galactic fountain and feedback models.

%---------------------------------
%\begin{acknowledgments}
%\end{acknowledgments}

%% To help institutions obtain information on the effectiveness of their 
%% telescopes the AAS Journals has created a group of keywords for telescope 
%% facilities.
%
%% Following the acknowledgments section, use the following syntax and the
%% \facility{} or \facilities{} macros to list the keywords of facilities used 
%% in the research for the paper.  Each keyword is check against the master 
%% list during copy editing.  Individual instruments can be provided in 
%% parentheses, after the keyword, but they are not verified.

\begin{acknowledgments}
    We are very grateful to the anonymous referee for the constructive, thoughtful, and helpful comments and suggestions on this paper. 
    BC would like to thank Yakov Faerman, Mario Juri\'c, Emily Levesque, and Matt McQuinn for their helpful comments on this manuscript, as well as Gabriele Ponti for valuable conversations. JKW would like to thank Chris Howk, Nicolas Lehner, Andy Fox and Jonathan Stern for helpful conversations around the implementation of this method and the interpretation of our results. Support for this work was provided by NSF-CAREER 2044303. This research was supported by the Munich Institute for Astro-, Particle and BioPhysics (MIAPbP) which is funded by the Deutsche Forschungsgemeinschaft (DFG, German Research Foundation) under Germany´s Excellence Strategy – EXC-2094 – 390783311. 
    Some of the data presented in this paper were obtained from the Mikulski Archive for Space Telescopes (MAST) at the Space Telescope Science Institute. The specific observations analyzed can be accessed via \dataset[https://doi.org/10.17909/k15n-zp34]{https://doi.org/10.17909/k15n-zp34}. STScI is operated by the Association of Universities for Research in Astronomy, Inc., under NASA contract NAS5–26555. Support to MAST for these data is provided by the NASA Office of Space Science via grant NAG5–7584 and by other grants and contracts. 
    This research has made use of the HSLA database, developed and maintained at STScI, Baltimore, USA, and data from the Wisconsin H$\alpha$ Mapper  sky survey (WHAM-SS). WHAM-SS was funded primarily by the National Science Foundation, and the WHAM facility was designed and built with the help of the University of Wisconsin Graduate School, Physical Sciences Lab, and Space Astronomy Lab. NOAO staff at Kitt Peak and Cerro Tololo provided on-site support for its remote operation. 
\end{acknowledgments}

\vspace{5mm}
\facilities{HST(COS), HST(STIS), FUSE}

%% Similar to \facility{}, there is the optional \software command to allow 
%% authors a place to specify which programs were used during the creation of 
%% the manuscript. Authors should list each code and include either a
%% citation or url to the code inside ()s when available.

\software{\texttt{astropy} \citep{2013A&A...558A..33A,2018AJ....156..123A}, 
          CLOUDY \citep{ferland17}, 
          \texttt{JAX} \citep{jax2018github}, 
          \texttt{linetools} \citep{prochaska17a}, 
          \texttt{pyigm} \citep{pyigm}, \texttt{veeper} \citep{burchett24}
          }

%% Appendix material should be preceded with a single \appendix command.
%% There should be a \section command for each appendix. Mark appendix
%% subsections with the same markup you use in the main body of the paper.

%% Each Appendix (indicated with \section) will be lettered A, B, C, etc.
%% The equation counter will reset when it encounters the \appendix
%% command and will number appendix equations (A1), (A2), etc. The
%% Figure and Table counter will not reset.

\appendix

\section{Ion Column Density Measurements and the Voigt Profile Fitting Parameters} \label{sec:appendix}

We present all the electron column density $N_e$ measured by pulsar DM and ion column density measurements $N_{ion}$ for every GC sight line studied in this paper. The discussion about 
an upper limit and lower limit can be found in Section~\ref{sec:measurement}.
For the GCs with multiple available pulsar DMs, the mean value of the DMs is adopted and the uncertainty is given by 5 times of the standard error of the mean. 

%-----------------------------------------
\startlongtable
\begin{deluxetable*}{lccccc}
    \tablewidth{\textwidth}
    \tablecaption{Voigt profile measurements and ion column densities of the detected ionized gas clouds \label{tab:appendix_measure}}
    \tablehead{\colhead{GC} &\colhead{Component} & \colhead{Ion} & \colhead{$\log{N}$} & \colhead{$b$} & \colhead{$v_{\rm LSR}$} \\ 
    \colhead{} & \colhead{} & \colhead{} & \colhead{($\rm cm^{-2}$)} & \colhead{($\rm km\ s^{-1}$)} & \colhead{($\rm km\ s^{-1}$)} }
    \startdata
    M3 (NGC 5272) & Total & $e^{-}$ & $<$ \tablenotemark{a} 19.91 $\pm$ 0.01 & \dotfill & \dotfill \\
            & IVa & \ion{H}{1}  & $>$ 16.96 $\pm$ 0.16 & \dotfill & \dotfill \\
            &     & \ion{O}{1}  & $>$ 14.81 $\pm$ 0.07 & 21.8 $\pm$ 5.3 & -38.3 $\pm$ 2.7 \\
            &     & \ion{Fe}{2} & 14.92 $\pm$ 0.02 & 27.9 $\pm$ 0.8 & -37.5 $\pm$ 0.7 \\
            &     & \ion{S}{2}  & 14.83 $\pm$ 0.04 & 12.5 $\pm$ 1.9 & -38.0 $\pm$ 1.7 \\
            &     & \ion{Fe}{3} & 14.40 $\pm$ 0.08 & 19.8 $\pm$ 2.9 & -45.8 $\pm$ 1.7 \\
            &     & \ion{S}{3}  & 14.41 $\pm$ 0.10 & 21.9 $\pm$ 3.5 & -23.7 $\pm$ 4.1 \\
            &     & \ion{Si}{4} & 13.63 $\pm$ 0.11 & 16.9 $\pm$ 3.1 & -38.2 $\pm$ 1.8 \\
            &     & \ion{C}{4}  & 14.21 $\pm$ 0.06 & 19.6 $\pm$ 3.5 & -39.6 $\pm$ 1.7 \\
            &     & \ion{N}{5}  & 13.15 $\pm$ 0.05 & 19.1 $\pm$ 4.3 & -33.2 $\pm$ 2.6\\
            & IVb & \ion{H}{1}  & $>$ 16.92 $\pm$ 0.16 & \dotfill & \dotfill \\
            &     & \ion{O}{1}  & 13.92 $\pm$ 0.29 & 8.6 $\pm$ 5.7 & -73.1 $\pm$ 4.1 \\
            &     & \ion{Fe}{2} & 14.02 $\pm$ 0.12 & 4.0 $\pm$ 0.9 & -73.3 $\pm$ 1.0 \\
            &     & \ion{Si}{2} & 14.23 $\pm$ 0.06 & 26.2 $\pm$ 2.7 & -76.7 $\pm$ 3.2 \\
            &     & \ion{S}{2}  & $<$ 14.63 $\pm$ 0.03 & \dotfill & \dotfill \\
            &     & \ion{C}{2}  & $>$ 14.64 $\pm$ 0.22 & 27.0 $\pm$ 3.4 & -61.9 $\pm$ 7.5 \\
            &     & \ion{Fe}{3} & 13.82 $\pm$ 0.23 & 21.9 $\pm$ 7.9 & -81.4 $\pm$ 8.9 \\
            &     & \ion{Si}{3} & $>$ 13.61 $\pm$ 0.28 & 25.2 $\pm$ 5.3 & -68.4 $\pm$ 11.0 \\
            &     & \ion{Si}{4} & 13.17 $\pm$ 0.30 & 28.0 $\pm$ 13.0 & -74.6 $\pm$ 15.6 \\
            &     & \ion{C}{4}  & 13.64 $\pm$ 0.14 & 17.8 $\pm$ 4.9 & -79.7 $\pm$ 4.9\\
            &     & \ion{N}{5}  & $<$ 12.55 $\pm$ 0.19 & \dotfill & \dotfill \\
    \hline
    M5 (NGC 5904) & Total & $e^{-}$ & $<$ \tablenotemark{b} 19.96 $\pm$ 0.02 & \dotfill & \dotfill \\
            & HVa & \ion{H}{1}  & $>$ 16.27 $\pm$ 0.09 & \dotfill & \dotfill \\
            &     & \ion{O}{1}  & 13.32 $\pm$ 0.04 & \dotfill & -125.8 \\
            &     & \ion{Fe}{2} & $<$ 13.25 $\pm$ 0.16 & \dotfill & \dotfill \\ 
            &     & \ion{Si}{2} & 12.88 $\pm$ 0.03 & 10.7 $\pm$ 0.9 & -124.2 $\pm$ 0.8 \\
            &     & \ion{S}{2}  & $<$ 13.26 $\pm$ 0.54 & \dotfill & \dotfill \\ 
            &     & \ion{C}{2}  & 14.11 $\pm$ 0.03 & 16.2 $\pm$ 0.9 & -125.5 $\pm$ 1.1 \\ 
            &     & \ion{Si}{3} & 13.14 $\pm$ 0.19 & 15.1 $\pm$ 4.6 & -115.2 $\pm$ 2.5 \\
            &     & \ion{Si}{4} & 12.55 $\pm$ 0.03 & 11.9 $\pm$ 1.4 & -111.4 $\pm$ 0.9 \\
            &     & \ion{C}{4}  & 13.55 $\pm$ 0.20 & 12.0 $\pm$ 3.0 & -113.4 $\pm$ 1.7 \\
            & HVb & \ion{H}{1}  & $>$ 16.00 $\pm$ 0.12 & \dotfill & \dotfill \\
            &     & \ion{O}{1}  & 13.31 $\pm$ 0.04 & \dotfill & -140.8 \\
            &     & \ion{Fe}{2} & $<$ 13.23 $\pm$ 0.18 & \dotfill & \dotfill \\
            &     & \ion{Si}{2} & 13.25 $\pm$ 0.02 & 5.7 $\pm$ 0.3 & -144.6 $\pm$ 0.3 \\
            &     & \ion{S}{2}  & $<$ 13.25 $\pm$ 0.34 & \dotfill & \dotfill \\
            &     & \ion{C}{2}  & 14.28 $\pm$ 0.49 & 4.4 $\pm$ 1.5 & -143.6 $\pm$ 0.5 \\
            &     & \ion{Si}{3} & 12.96 $\pm$ 0.09 & 6.9 $\pm$ 1.1 & -140.5 $\pm$ 1.0 \\
            &     & \ion{Si}{4} & 12.64 $\pm$ 0.02 & 6.9 $\pm$ 0.7 & -144.4 $\pm$ 0.4 \\
            &     & \ion{C}{4}  & 13.34 $\pm$ 0.05 & 8.7 $\pm$ 1.5 & -143.4  $\pm$ 1.0 \\
    \hline
    M10 (NGC 6254) & Total & $e^{-}$ & $<$ \tablenotemark{c} 20.13 $\pm$ 0.02 & \dotfill & \dotfill \\
            & IVa & \ion{O}{1}  & 14.68 $\pm$ 0.02 & 20.4 $\pm$ 1.2 & -62.9 $\pm$ 1.0 \\
            &     & \ion{Fe}{2} & 14.28 $\pm$ 0.01 & 25.0 $\pm$ 1.2 & -64.8 $\pm$ 0.7 \\ 
            &     & \ion{Si}{2} & 14.30 $\pm$ 0.01 & 16.2 $\pm$ 0.4 & -63.2 $\pm$ 0.5 \\
            &     & \ion{S}{2}  & 14.38 $\pm$ 0.03 & 22.0 $\pm$ 2.7 & -62.4 $\pm$ 1.5 \\
            &     & \ion{Si}{3} & $>$ 13.69 $\pm$ 0.07 & \dotfill & \dotfill \\
            &     & \ion{Si}{4} & 13.35 $\pm$ 0.02 & 28.1 $\pm$ 2.0 & -65.2 $\pm$ 1.3 \\
            &     & \ion{C}{4}  & 13.70 $\pm$ 0.03 & 26.7 $\pm$ 1.8 & -67.8 $\pm$ 1.5 \\
            &     & \ion{N}{5}  & $<$ 12.77 $\pm$ 0.36 & \dotfill & \dotfill \\
    \hline
    M13 (NGC 6205) & Total & $e^{-}$ & $<$ \tablenotemark{d} 19.97 $\pm$ 0.02 & \dotfill & \dotfill \\
            & HVa & \ion{O}{1}  & $>$ 14.64 $\pm$ 0.02 & 16.3 $\pm$ 0.9 & -95.0 $\pm$ 0.5 \\
            &     & \ion{Fe}{2} & 14.16 $\pm$ 0.01 & 21.4 $\pm$ 1.5 & -111.4 $\pm$ 0.7 \\
            &     & \ion{Si}{2} & 14.32 $\pm$ 0.01 & 22.9 $\pm$ 0.3 & -98.4 $\pm$ 0.4 \\
            &     & \ion{S}{2}  & 14.44 $\pm$ 0.01 & 13.3 $\pm$ 1.0 & -107.4 $\pm$ 0.6 \\
            &     & \ion{C}{2}  & $>$ 14.05 $\pm$ 0.07 & 29.2 $\pm$ 3.4 & -114.3 $\pm$ 9.2 \\
            &     & \ion{Fe}{3} & 14.17 $\pm$ 0.07 & 27.9 $\pm$ 3.9 & -96.1 $\pm$ 3.7 \\
            &     & \ion{S}{3} & 14.65 $\pm$ 0.02 & 23.6 $\pm$ 1.1 & -111.2 $\pm$ 0.9 \\
            &     & \ion{C}{4}  & 13.60 $\pm$ 0.42 & 18.5 $\pm$ 7.0 & -110.4 $\pm$ 12.0 \\
    \hline
    M15 (NGC 7078) & Total & $e^{-}$ & $<$ \tablenotemark{e} 20.32 $\pm$ 0.01 & \dotfill & \dotfill \\
            & IVa & \ion{H}{1}  & $>$ 16.86 $\pm$ 0.06 & \dotfill & \dotfill \\
            &     & \ion{O}{1}  & $>$ 14.86 $\pm$ 0.09 & 12.4 $\pm$ 1.8 & 62.0 $\pm$ 1.6 \\
            &     & \ion{Fe}{2} & 14.19 $\pm$ 0.02 & 17.9 $\pm$ 1.3 & 58.0 $\pm$ 0.8 \\ 
            &     & \ion{Si}{2} & 14.39 $\pm$ 0.02 & 16.1 $\pm$ 0.4 & 57.5 $\pm$ 0.5 \\
            &     & \ion{S}{2}  & 14.83 $\pm$ 0.02 & 10.7 $\pm$ 0.6 & 63.0 $\pm$ 0.4 \\
            &     & \ion{Fe}{3} & 14.02 $\pm$ 0.04 & \dotfill & \dotfill \\
            &     & \ion{Si}{4} & 13.14 $\pm$ 0.27 & 31.3 $\pm$ 11.9 & 42.6 $\pm$ 16.0 \\
            &     & \ion{C}{4}  & 13.27 $\pm$ 0.12 & 13.6 $\pm$ 4.6 & 61.8 $\pm$ 3.8 \\
    \hline
    NGC~6397     & Total & $e^{-}$ & $<$ \tablenotemark{f} 20.35 $\pm$ 0.01 & \dotfill & \dotfill \\
            & IVa & \ion{H}{1}  & $>$ 16.66 $\pm$ 0.10 & \dotfill & \dotfill \\
            &     & \ion{O}{1}  & 13.81 $\pm$ 0.02 & 7.7 $\pm$ 0.6 & 43.6 $\pm$ 0.4 \\
            &     & \ion{Fe}{2} & $<$ 12.84 $\pm$ 0.63 & \dotfill & \dotfill \\
            &     & \ion{S}{2}  & $<$ 13.98 $\pm$ 0.05 & \dotfill & \dotfill \\
            &     & \ion{C}{2}  & $>$ 14.12 $\pm$ 0.06 & 16.5 $\pm$ 1.3 & 59.5 $\pm$ 1.8 \\
            &     & \ion{Fe}{3} & 13.70 $\pm$ 0.13 & 23.0 $\pm$ 8.8 & 46.4 $\pm$ 5.3 \\
            &     & \ion{Si}{4} & 12.53 $\pm$ 0.05 & 14.4 $\pm$ 1.9 & 59.8 $\pm$ 1.3 \\
            &     & \ion{C}{4}  & 13.29 $\pm$ 0.12 & 16.0 $\pm$ 3.5 & 57.6 $\pm$ 3.7 \\
            &     & \ion{N}{5}  & 13.24 $\pm$ 0.38 & 27.3 $\pm$ 15.6 & 65.5 $\pm$ 15.7 \\
    \hline
    NGC~5139      & Total & $e^{-}$ & $<$ \tablenotemark{g} 20.48 $\pm$ 0.03 & \dotfill & \dotfill \\
            & IVa & \ion{N}{1}  & $>$ 15.07 $\pm$ 0.21 & 14.6 $\pm$ 2.1 & -39.6 $\pm$ 4.7 \\
            &     & \ion{O}{1}  & $>$ 14.77 $\pm$ 0.27 & 15.0 $\pm$ 4.8 & -48.2 $\pm$ 5.9 \\
            &     & \ion{Fe}{2} & 15.20 $\pm$ 0.03 & 8.6 $\pm$ 0.4 & -50.7 $\pm$ 0.5 \\
            &     & \ion{Ni}{2} & 13.91 $\pm$ 0.07 & 14.2 $\pm$ 1.8 & -43.8 $\pm$ 1.8 \\
            &     & \ion{S}{2}  & 15.53 $\pm$ 0.08 & 20.8 $\pm$ 1.8 & -34.5 $\pm$ 3.1 \\ 
            &     & \ion{Si}{4} & 13.57 $\pm$ 0.18 & 12.0 $\pm$ 3.4 & -44.3 $\pm$ 1.8 \\
            &     & \ion{C}{4}  & 13.99 $\pm$ 0.19 & 18.5 $\pm$ 4.8 & -32.4 $\pm$ 2.3 \\
            &     & \ion{N}{5}  & 13.39 $\pm$ 0.15 & 25.0 $\pm$ 8.5 & -45.6 $\pm$ 7.2 \\
    \hline
    NGC~6752     & Total & $e^{-}$ & $<$ \tablenotemark{h} 20.01 $\pm$ 0.01 & \dotfill & \dotfill \\
            & IVa & \ion{O}{1}  & 15.15 $\pm$ 0.04 & 14.2 $\pm$ 1.4 & 38.8 $\pm$ 1.2 \\
            &     & \ion{Fe}{2} & 14.15 $\pm$ 0.04 & 14.5 $\pm$ 1.8 & 45.7 $\pm$ 1.7 \\
            &     & \ion{Si}{2} & 14.61 $\pm$ 0.37 & 19.6 $\pm$ 7.8 & 39.6 $\pm$ 11.1 \\
            &     & \ion{S}{2}  & 14.59 $\pm$ 0.08 & 18.4 $\pm$ 3.1 & 41.5 $\pm$ 3.3 \\
            &     & \ion{Fe}{3} & 13.73 $\pm$ 0.09 & 12.7 $\pm$ 3.4 & 55.2 $\pm$ 2.2 \\
            &     & \ion{Si}{4} & 13.29 $\pm$ 0.04 & 28.0 $\pm$ 2.8 & 47.7 $\pm$ 2.3 \\
            &     & \ion{C}{4}  & 13.86 $\pm$ 0.05 & 20.1 $\pm$ 2.5 & 45.3 $\pm$ 2.4 \\
            &     & \ion{N}{5}  & 13.37 $\pm$ 0.32 & 33.6 $\pm$ 17.8 & 45.7 $\pm$ 20.4 \\
    \enddata
    \tablerefs{(a) \cite{hessels07}; (b) \cite{freire08, pallanca14}; (c) \cite{pan21}; (d) \cite{wang20}; (e) \cite{anderson93}; (f) \cite{damico01, zhang22}; (g) \cite{dai20}; (h) \cite{damico02}}
    \tablecomments{The electron column density given by pulsar DM is the total column density for the GC sight line. $\log{N}$ is ion column density, and $b$ is the Doppler parameter of the absorption line. The errors correspond to 1$\sigma$ given by Voigt profile fitting procedure using \texttt{veeper}.}
\end{deluxetable*}
%\section{}
%------------------------------

\section{Collisional ionization} \label{subsubsec:collisional}

While the two-phase photoionization model shows a good agreement with the observation for \ion{Si}{4} and \ion{C}{4} (Fig.~\ref{fig:abundance_deviation}), the observed \ion{N}{5} column densities still show about 1~dex excess compared to the model, supporting the idea that photoionization alone is insufficient to produce ions with such a high ionization state \citep{fox04}. 
Consequently, the intermediate ions potentially reside in a hotter gas phase ($T \gtrsim 10^5~K$) where collisional ionization becomes dominant \citep{fox05, savage09, werk19}.

At the DHI, a gas phase at transitional temperature of $10^5$~K can arise from various collisional ionization mechanisms and processes, encompassing collisional ionization equilibrium (CIE; \citealt{sutherland93}), radiative cooling of hot gas \citep{edgar86}, turbulent mixing layers \citep{slavin93, fielding20, tan21}, conduction fronts \citep{borkowski90}, and shock ionization \citep{dopita96}. 
We evaluate multiple potential mechanisms against the data, specifically examining their compatibility with the photoionization model presented in Section~\ref{subsec:low ion result} to ensure robust consistency. 
While a detailed analysis of absorption line profiles and column density ratios could serve as diagnostics for ionization mechanisms and gas properties \citep{fox04, gnat07}, such an examination is beyond the scope of this work, which primarily centers on deriving the metallicity of ionized gas. 
Additionally, our observed set of ions is limited for such analysis, predominantly composed of low ions and occasionally missing higher ions than \ion{C}{4}.

%%-------------------
\begin{figure}
    \centering
    \includegraphics[width=0.44\textwidth]{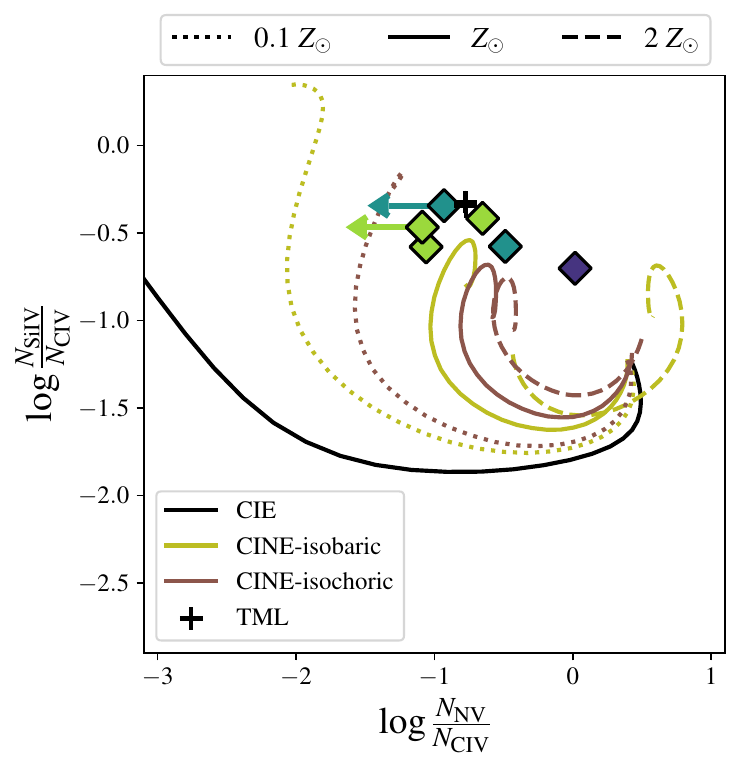}
    \caption{Column density ratios of \ion{N}{5}/\ion{C}{4} vs. \ion{Si}{4}/\ion{C}{4}, comparing the observation and predicted values by various ionization mechanisms, CIE (black), isobaric (olive) and isochoric (brown) cooling, and turbulent mixing layers (plus mark). Non-equilibrium cooling depends on metallicity as the model of 0.1 (dotted), 1 (solid), and $2\ Z_{\odot}$ (dashed) shows.}
    \label{fig:ionization_mechanism}
\end{figure}
%-----------------------

We compare the observed intermediate ions with CIE, non-equilibrium radiative cooling, and turbulent mixing layers models among various mechanisms. 
Other mechanisms are excluded because of their complexity and dependence on uncertain parameters. 
For instance, thermal conduction is intricately linked to magnetic field \citep{borkowski90}. 
Shock ionization typically results from fast shocks \citep{dopita96}, however, the detected clouds are mostly IVCs that exhibit a relatively low velocity ($|v_{\rm LSR}| < 90\ \rm km\ s^{-1}$, though this is a projected velocity), making them less likely to be relevant.

CIE serves as a basic depiction of collisionally ionized gas, assuming a balance between ionization and recombination at a fixed temperature. 
When the timescale for radiative cooling is rapid compared to recombination, gas deviates from CIE, necessitating consideration of non-equilibrium radiative cooling. 
We adopt the CIE and non-equilibrium cooling models developed by \cite{gnat07} for a temperature range of $10^{4.5} \leq T \leq 10^{5.5}$~K. 
Isobaric (constant pressure) and isochoric (constant density) cooling conditions are explored for gas with metallicity of $Z/Z_{\odot} = 0.1, 1$, and 2, reflecting our wide-ranging metallicity measurements. 
Turbulent mixing layers are relevant to DHI gas, given their prevalence in multiphase gas and ample multiphase gas flows at the DHI. 
These layers can be conceptualized in scenarios involving neutral clouds either ejected from the disk via feedback \citep{fraternali17} or accreted from outside of the Milky Way \citep{heitsch09} and traveling through a hot halo medium. 
We apply the analytic model of turbulent radiative mixing layers developed by \cite{chen23} with $p_{\rm hot}/k_{\rm B} = 10^3\ \rm K\ cm^{-3}$, considering a hot medium with a temperature of $T = 10^{6}$~K and hydrogen number density of $n_{\rm H} = 10^{-3}\ \rm cm^{-3}$, which correspond to those of hot halo gas. 
We note that the prediction from turbulent radiative mixing layers shows almost constant column density ratios for $p_{\rm hot}/k_{\rm B} = 10^{3-5}\ \rm K\ cm^{-3}$.

We utilize column density ratios, specifically $N_{\rm Si IV}/N_{\rm C IV}$ and $N_{\rm N V}/N_{\rm C IV}$, as diagnostic tools for ionization mechanisms, displayed in Fig.~\ref{fig:ionization_mechanism}. 
The observational constraints include instances where \ion{N}{5} absorption was detected or where no interface from other absorption features hindered the estimation of the upper limit of \ion{N}{5}, denoted by diamond points. 
The color represents metallicity derived from low ions (Section~\ref{subsec:low ion result}) categorized into sub-solar ($< 0.3\ Z_{\odot}$; blue), near-solar ($0.3-1\ Z_{\odot}$; purple), and super-solar ($> Z_{\odot}$; pink). 
The observed $\log{N_{\rm SiIV}/N_{\rm CIV}}$ spans a range between -1 and -0.35, significantly surpassing the predicted ratio by CIE (black solid line). 
As a result, CIE is ruled out as an explanation for the highly ionized gas portion of the detected DHI clouds. 
The observed column density ratios fall within the isobaric (olive) and isochoric (brown) cooling regimes, while the models show considerable dependence on metallicity. 
We note that the metallicity estimate based on low ions may differ from the transitional temperature gas, especially in situations where metal-poor or metal-rich cool gas travels through an ambient hot halo medium. This difference arises because low ions primarily trace the colder core of the cloud, whereas the transitional gas could represent a more diffuse wake produced by the cooling of halo gas.  
The turbulent radiative mixing layers model with solar abundance (plus mark) also matches with the observed ion ratios. 

%%--------------------
\begin{figure}
    \centering
    \includegraphics[width=0.44\textwidth]{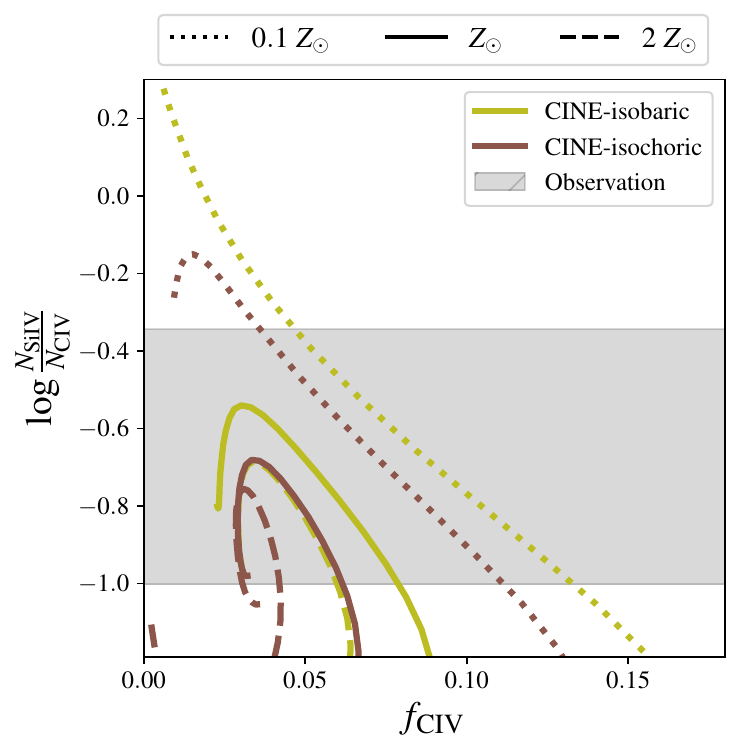}
    \caption{\ion{C}{4} fraction to total carbon and column density ratios of \ion{Si}{4} and \ion{C}{4} predicted by isobaric (olive) and isochoric (brown) non-equilibrium radiative cooling at 0.1, 1, and $2\ Z_{\odot}$ metallicity. The observed column density ratios of \ion{Si}{4} and \ion{C}{4} are within the gray-shaded range.}
    \label{fig:f_civ}
\end{figure}
%%---------------------

To corroborate the consistency of collisional ionization models with the photoionization model, we calculate the fraction of gas in the transitional temperature phase to the photoionized low-ion gas ($T = 10^{3.5-4}$~K). 
For the photoionized gas, we use the ionized hydrogen column density provided in Table~\ref{tab:best-fit}. 
As shown in Fig.~\ref{fig:f_civ}, we employ \ion{C}{4} fraction to the total carbon, $f_{\rm CIV}$, to estimate the ionized hydrogen column density of the transitional phase gas, $N_{\rm HII, trans}$. 
The \ion{C}{4} fraction, determined by the models using the column density ratio of \ion{Si}{4} and \ion{C}{4}, varies with metallicity. 
Assuming the cooling of hot halo gas at sub-solar metallicity ($Z \sim 0.1\ Z_{\odot}$; \citealt{ponti23}), we find $N_{\rm HII, trans}$ to be $25-150~\%$ of the photoionized gas, and these values remain consistent within the upper limit of the total electron column density provided by pulsar DMs along the sight line. 
On the other hand, for solar-metallicity hot gas expected in feedback-driven outflows, the model fails to find a solution for gas clouds with a high \ion{Si}{4}/\ion{C}{4} ratio (Fig.~\ref{fig:f_civ}), and it yields much lower $N_{\rm HII, trans}$ ranging from $4-30~\%$ of the photoionized gas. 

The observed intermediate ions, \ion{Si}{4}, \ion{C}{4}, and \ion{N}{5}, are more likely to reside in a gas phase at transitional temperature of $10^5$~K, rather than in photoionized gas ($T \sim 10^4$~K). This temperature regime can be attributed to radiative cooling of hot gas and turbulent mixing layers, along with the observed ion ratios shown in Fig.~\ref{fig:ionization_mechanism}. 
Additionally, our photoionization model for low ions is compatible with collisional ionization for intermediate ions under moderate condition, maintaining consistency with electron column densities derived from pulsar DMs.

%% For this sample we use BibTeX plus aasjournals.bst to generate the
%% the bibliography. The sample631.bib file was populated from ADS. To
%% get the citations to show in the compiled file do the following:
%%
%% pdflatex sample631.tex
%% bibtext sample631
%% pdflatex sample631.tex
%% pdflatex sample631.tex

\bibliography{references}{}
\bibliographystyle{aasjournal}

%% This command is needed to show the entire author+affiliation list when
%% the collaboration and author truncation commands are used.  It has to
%% go at the end of the manuscript.
%\allauthors

%% Include this line if you are using the \added, \replaced, \deleted
%% commands to see a summary list of all changes at the end of the article.
%\listofchanges

\end{document}